\documentclass[journal,10pt,twoside]{IEEEtran}

\usepackage{ifpdf}
\usepackage{graphicx}
\usepackage{cite}
\usepackage[cmex10]{amsmath}
\usepackage{amssymb}
\usepackage{algorithmic}
\usepackage{array}
\usepackage{color}
\usepackage{colortbl}
\usepackage[table]{xcolor}
\usepackage{mdwmath}
\usepackage{mdwtab}
\usepackage{url}
\usepackage{epsfig}
\usepackage{float}
\usepackage{balance}
\usepackage{lettrine}
\usepackage{empheq}
\usepackage{color}
\usepackage{epstopdf}
\usepackage{algorithm}
\usepackage{textcomp}
\usepackage{multirow}
\usepackage{subcaption}
\usepackage{dblfloatfix}

\usepackage{acronym}

\begin{document}
\title{Hybrid LiFi and WiFi Networks: A Survey}
\author{Xiping Wu\thanks{X. Wu and L. Zhou are with the Department of Engineering Science, University of Oxford, Oxford, OX1 3PJ, UK, e-mail:\{xiping.wu, lai.zhou\}@eng.ox.ac.uk. \newline  
\indent M. D. Soltani, M. Safari and H. Haas are with the LiFi Research and Development Centre, School of Engineering, The University of Edinburgh, Edinburgh, EH9 3JL, UK, e-mail:\{m.dehghani, majid.safari, h.haas\}@ed.ac.uk}, \emph{Member, IEEE}, Mohammad Dehghani Soltani, \emph{Student Member, IEEE}, Lai Zhou, Majid Safari, \emph{Senior Member, IEEE}, and Harald Haas, \emph{Fellow, IEEE}
}
\maketitle

\begin{abstract}

To tackle the rapidly growing number of mobile devices and their expanding demands for Internet services, network convergence is envisaged to integrate different technology domains. A recently proposed and promising approach to indoor wireless communications is integrating light fidelity (LiFi) and wireless fidelity (WiFi), namely a hybrid LiFi and WiFi network (HLWNet). This type of network combines the high-speed data transmission of LiFi and the ubiquitous coverage of WiFi. In this paper, we present a survey-style introduction to HLWNets, starting with a framework including the network structure, cell deployment, multiple access schemes, modulation techniques, illumination requirements and backhauling. Then, key performance metrics and recent achievements are reviewed. Further, the unique challenges faced by HLWNets are elaborated in many research directions, including user behavior modeling, interference management, handover and load balancing. Finally, we discuss the potential of HLWNets in application areas such as indoor positioning and physical layer security.


\end{abstract}


\section{Introduction}

\lettrine[loversize=0.1, nindent=0em]{T}{HE} recent visual networking index published by Cisco systems predicts that by 2022, mobile data traffic will account for 71 percent of Internet protocol traffic, and more than 80\% of mobile data traffic will occur indoors \cite{VNI_2022}. This drives short-range wireless communication technologies such as wireless fidelity (WiFi) to become a key component in the fifth generation and beyond (5GB) era. Globally, there will be nearly 549 million public WiFi hotspots by 2022, up from 124 million hotspots in 2017 \cite{VNI_2022}. Due to the limited spectrum resource of radio frequency (RF), the dense deployment of WiFi hotspots would result in intense competitions for available channels. This challenges the RF system to meet the exponentially increasing demand for mobile data traffic, which will increase seven-fold between 2017 and 2022, reaching 77.5 exabytes per month by 2022.

In order to tackle the looming spectrum shortage in RF, wireless communication technologies employing extremely high frequencies have drawn significant attentions. Among these technologies is light fidelity (LiFi) \cite{7360112}. Using light waves as a signal bearer, this relatively new technology is able to exploit the vast optical spectrum, nearly 300 THz. LiFi access points (APs) can be integrated into the existing lighting infrastructure, e.g. light-emitting diode (LED) lamps, realizing a dual purpose system which provides illumination and communication. Recent research demonstrates that with a single LED, LiFi is capable of achieving peak data rates above 10 Gbps \cite{10gbps}. LiFi offers many other advantages over its RF counterpart, including: i) a licence-free optical spectrum; ii) the ability to be used in RF-restricted areas, e.g. hospitals and underwater; and iii) the capability of providing secure wireless communications, as light does not penetrate opaque objects. Also, LiFi has some limitations as it: i) covers a relatively short range, usually a few meters with a single AP; and ii) is susceptive to connectivity loss due to obstructions. Nevertheless, as a complementary approach to WiFi, LiFi is a promising technology to fulfill the future demand for data rates.

\begin{figure}[!b]
\centering
\includegraphics[width=3.2in]{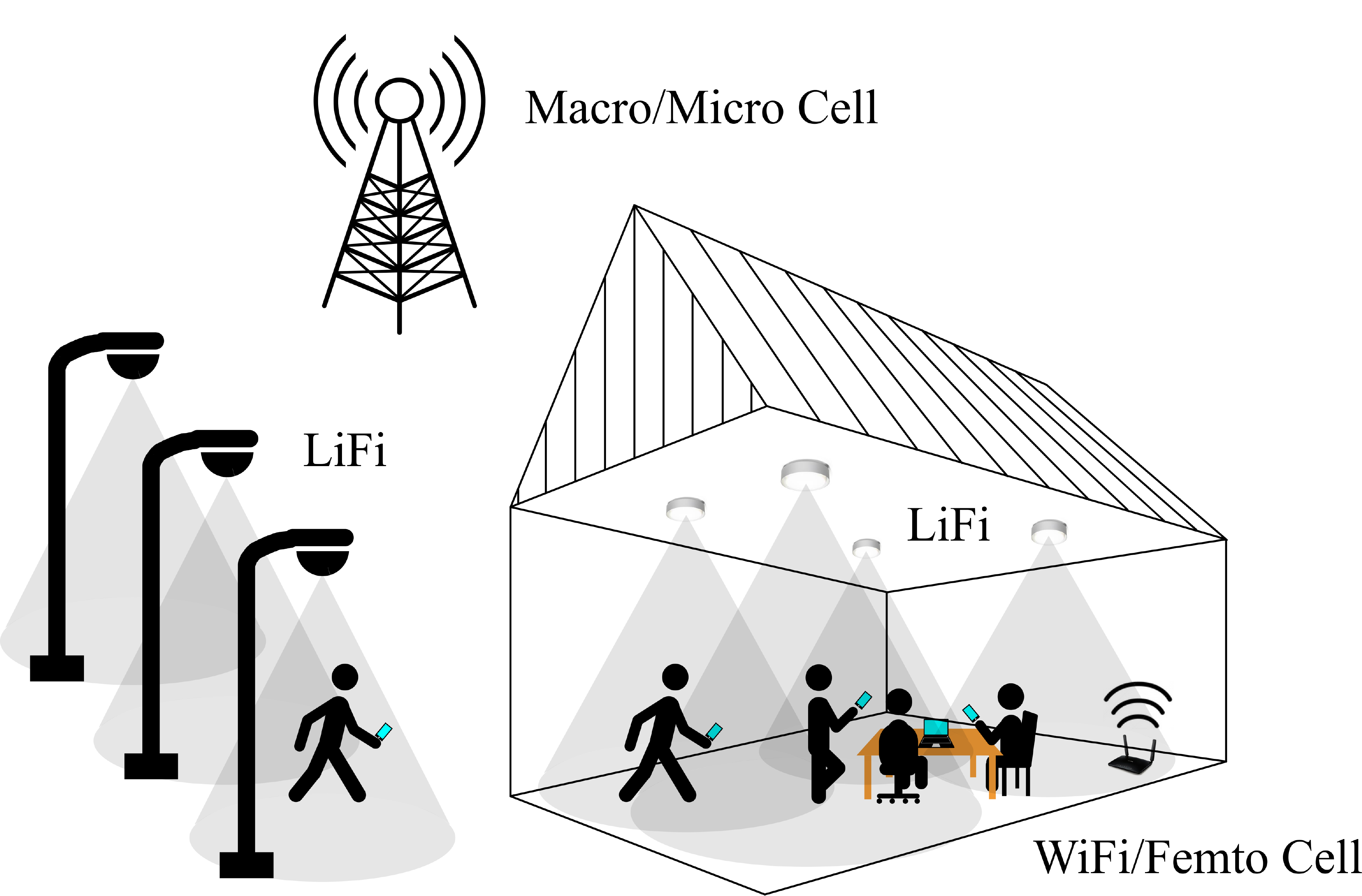}
\caption{Network convergence in 5G and beyond.}
\label{fig:LiFi_in_5G}
\end{figure}

Combining the high-speed data transmission of LiFi and the ubiquitous coverage of WiFi, the concept of a hybrid LiFi and WiFi network (HLWNet) was first mentioned by Rahaim \emph{et al.} in 2011 \cite{6162563}. Soon later, Stefan \emph{et al.} \cite{6965999} extended the research to integrate LiFi and femtocells. This type of hybrid network is proven to achieve a better network performance than a stand-alone LiFi or RF system \cite{7145863}. Fig. \ref{fig:LiFi_in_5G} presents a vision of integrating mainstream wireless networks and LiFi in 5GB environments. In outdoor scenarios, users can be served by macro/micro cells or LiFi-enabled street lamps. When users move indoor, they can be shifted to a HLWNet for higher quality of service.

\begin{table*}[t]
\renewcommand{\arraystretch}{1.2}
\caption{List of Acronyms}
\label{table:acronyms}
\centering
\begin{tabular}{|l|l|l|l|}
\hline
Acronym & Description                                    & Acronym & Description     \\                           
\hline
ACK & Acknowledgment                                     & LB & Load Balancing \\
ACO-OFDM & Asymmetrically clipped optical OFDM           & LED & Light-Emitting Diode \\
ADR & Angle Diversity Receiver                           & LiFi & Light Fidelity \\
AOA & Angle of Arrival                                   & LoS & Line of sight\\
AP & Access Point                                        & LTE & Long-term evolution\\
ASE & Area spectral efficiency                           & MAC & Medium Access Control \\
BD & Block Diagonalization                               & MIMO & Multiple-Input Multiple-Output \\
BER & Bit-error ratio                                    & MUD & Multi-User Detection \\
BIA &  Blind Interference Alignment                      & NOMA & Non-orthogonal multiple access\\
CCI & Co-Channel Interference                            & OFDM & Orthogonal frequency-division multiplexing\\
CDMA & Code-Division Multiple Access                     & OFDMA & Orthogonal frequency-division multiple access\\
CoMP & Coordinated Multipoint                            & O-OFDM & Optical-OFDM\\
CP & Cyclic prefix                                       & ORWP & Orientation-based random waypoint \\
CSI & Channel State Information                          & PD & Photo Diode \\
CSIT & Channel State Information at the Transmitter      & PLC & Power-line communication \\
CSMA/CA & Carrier sense multiple access with collision avoidance        & PLS & Physical Layer Security \\
CTS & Clear-to-send                                      & POE & Power over Ethernet\\
DAC & Digital to analog converter                        & PPP & Poisson point process\\
DC & Direct current                                      & QoS & Quality of service\\
DCO-OFDM & Direct current-biased optical OFDM            & RF & Radio Frequency \\
DIFS & Distributed inter-frame space                     & RSS & Received Signal Strength \\
FFR & Fractional Frequency Reuse                         & RTS & Request-to-send\\
FFT & Fast Fourier transform                             & RWP & Random waypoint \\
FoV & Field of view                                      & SDMA & Space-division multiple access\\
FR & Frequency Reuse                                     & SDN & Software-Defined Network \\
GoF & Grade of fairness                                  & SFR & Soft Frequency Reuse \\ 
GPS & Global Positioning System                          & SIFS & Short inter-frame space\\
HetNet & Heterogeneous Networks                          & SINR & Signal-to-interference-plus-noise ratio\\
HHO & Horizontal Handover                                & SOTA & State-of-the-Art \\
HLWNet & Hybrid LiFi and WiFi Network                    & TDMA & Time-division multiple access\\
HS & Handover Skipping                                   & TDOA & Time Difference of Arrival \\
ICI & Inter-channel interference                         & TOA & Time of Arrival \\
IFFT & Inverse fast Fourier transform                    & VHO & Vertical Handover \\
IM/DD & Intensity modulation and direct detection        & VLC & Visible Light Communication \\
IPS & Indoor Positioning System                          & WiFi & Wireless Fidelity \\
ISI & Inter-symbol interference                          & ZF & Zero Forcing \\
\hline
\end{tabular}
\end{table*}

The state-of-the-art (SOTA) and research challenges of LiFi were summarized in multiple articles \cite{6191306,7072557,7239528}, including the fields of system design, optical multiple-input multiple-output (MIMO), channel coding and networking. However, to the best of our knowledge, no survey work has so far been conducted for HLWNets, where seamless integration and intelligent resource allocation between different technology domains become paramount. In \cite{7293077}, the system design of HLWNets was investigated and the network performance was analyzed. Although this research involves a simple model of light-path blockage, the optimal traffic allocation is only developed for stationary users. Apart from that, \cite{7293077} fails to consider the frequency reuse among LiFi APs and the consequent interference management. A similar system design work was carried out in \cite{7921566}, but with the functionality of spectrum and power allocation. Nonetheless, this work still targets stationary users and lacks a comprehensive overview of the HLWNet.

The main contributions of this paper are reviewing the SOTA of HLWNets, addressing the unique challenges faced by HLWNets, and discussing the open issues and research directions. Specifically, the following technical challenges are mainly discussed: \textbf{1) system design}: the deployment of LiFi and WiFi APs is of vital importance to the network performance, while the LiFi setup also needs to meet illumination requirements; \textbf{2) resource allocation}: load balancing is critical for HLWNets since WiFi APs are susceptible to overload; \textbf{3) user mobility}: as HLWNet APs (especially LiFi) have a relatively short coverage range, user movement with an even moderate speed could cause frequent handovers. Also, LiFi channels are subject to light-path blockages, while varying receiver orientations could severely affect user association; \textbf{4) the benefit of HLWNets to application services}: HLWNets can not only improve the network performance in terms of throughput and latency, but also benefit application services such as indoor positioning and physical layer security.

The remainder of the paper is organized as follows. \mbox{Section \ref{sec:HLWNet}} presents the framework of HLWNets, including the network structure, cell deployment, multiple access, modulation techniques, illumination requirements and backhauling. Key performance metrics are summarized in \mbox{Section \ref{sec:performance_metrics}}. User behavior modeling is introduced in \mbox{Section \ref{sec:behavior_modeling}}, and the present works related to interference management are reviewed in \mbox{Section \ref{sec:interference_management}}. Handover and load balancing are elaborately discussed in \mbox{Section \ref{sec:handover}} and \ref{sec:load_balancing}, respectively. The advancements of HLWNets in indoor positioning and physical layer security are given in \mbox{Section \ref{sec:applications}}. Finally, conclusions are drawn in \mbox{Section \ref{sec:conclusion}}. The acronyms used in the paper are listed in \mbox{Table \ref{table:acronyms}}.


\section{A Framework of HLWNet} \label{sec:HLWNet}

In this section, we introduce a framework of HLWNet which consists of six components: i) network structure, ii) cell deployment, iii) multiple access schemes, iv) modulation techniques, v) illumination requirements, and vi) backhauling. This section aims to provide guidelines for designing a HLWNet system.

\subsection{Network Structure}

\begin{figure}[t]
\centering
\includegraphics[width=1\columnwidth,draft=false]{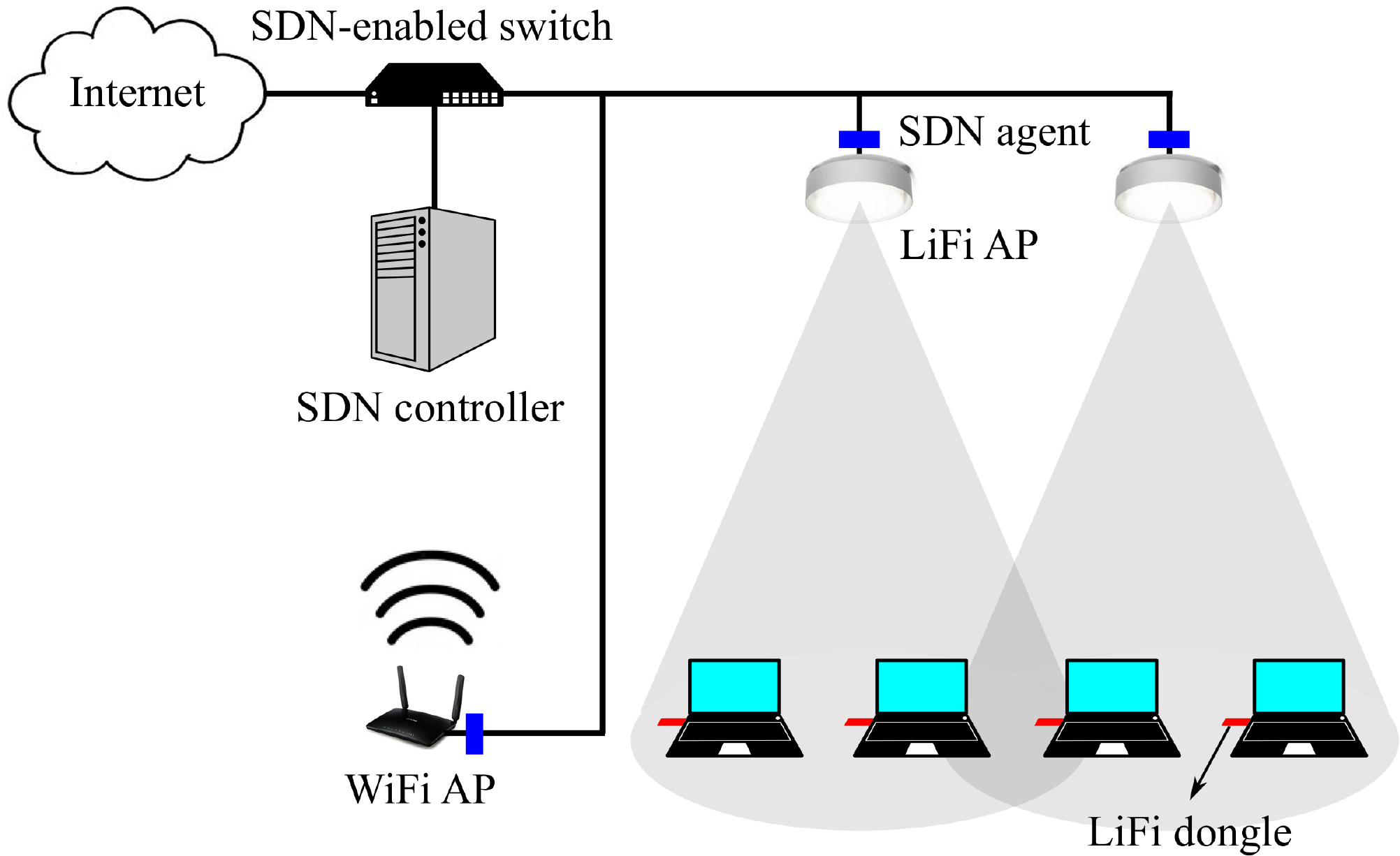}
\caption{Network structure of HLWNets based on an SDN platform.}
\label{fig:HLWNet_SDN_structure}
\vspace{-0.5cm}
\end{figure}

In order to achieve a flexible user association and resource allocation, a hypervisor managing all APs is essential. This can be realized by using software-defined networking (SDN) \cite{6994333}, which decouples the control plane from the data plane of forwarding devices. The development of SDN platforms for HLWNets is still in its infancy. In \cite{7784888}, an application-medium access control (MAC) cross-layer scheme was proposed, which employs flow admission control to dynamically allocate resources of LiFi and WiFi MAC layers. An SDN-enabled switch connects LiFi and WiFi APs, and extracts key performance indicator (KPI) information from them through SDN agents, as shown in Fig. \ref{fig:HLWNet_SDN_structure}. This information is then sent to an SDN controller, which makes decisions on the flow routes of each incoming data packet.

\subsection{Cell Deployment}
\begin{figure*}[t!]
		\centering
		
		\begin{subfigure}[b]{0.65\columnwidth}
			\centering
			\includegraphics[width=1\columnwidth,draft=false]{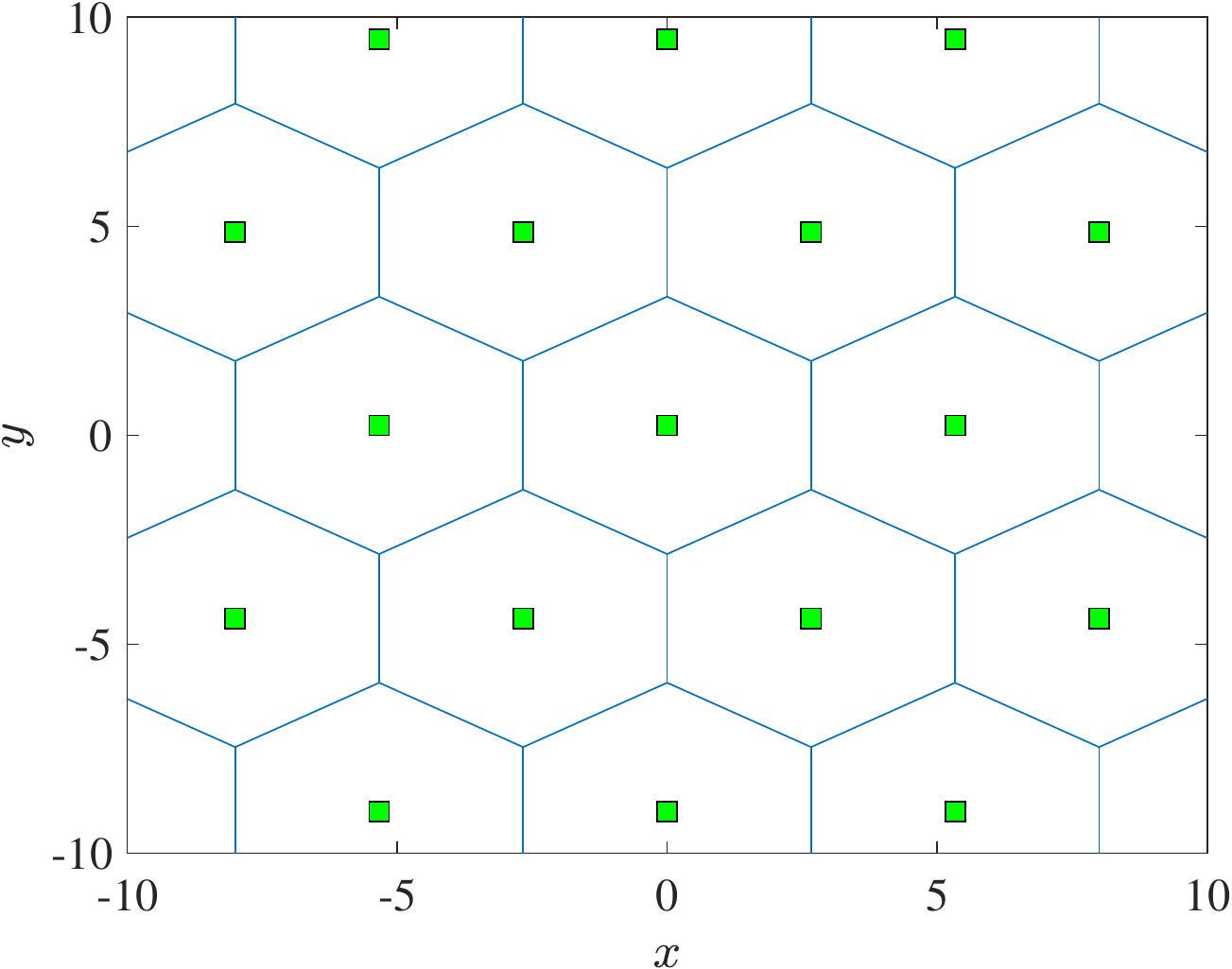}
						\caption{Hexagon}
				\label{Hex}
		\end{subfigure}%
		~
		\begin{subfigure}[b]{0.65\columnwidth} 
			\centering
			\includegraphics[width=1\columnwidth,draft=false]{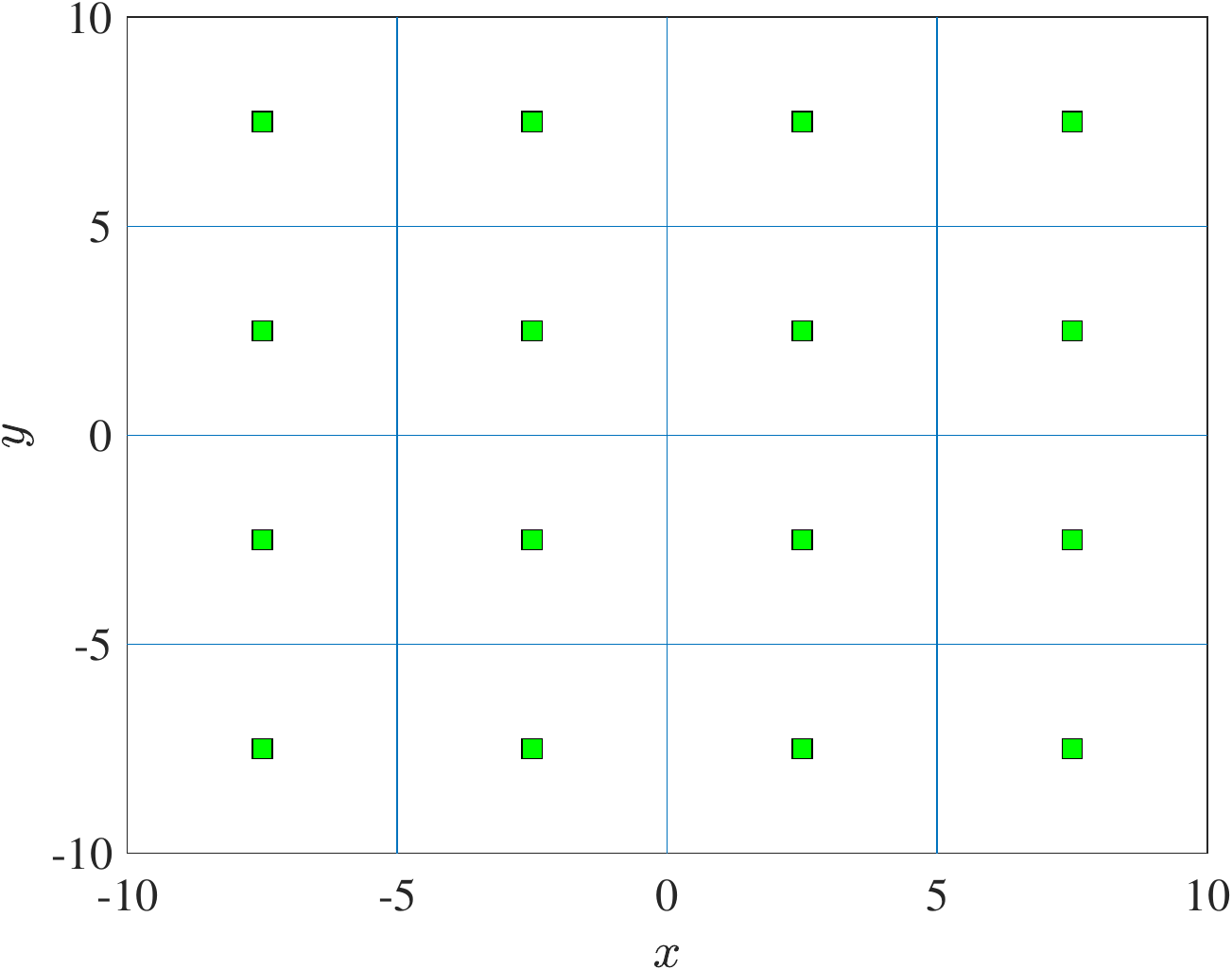}
						\caption{Lattice}
				\label{square}
		\end{subfigure}%
		~
		\begin{subfigure}[b]{0.65\columnwidth}
			\centering
			\includegraphics[width=1\columnwidth,draft=false]{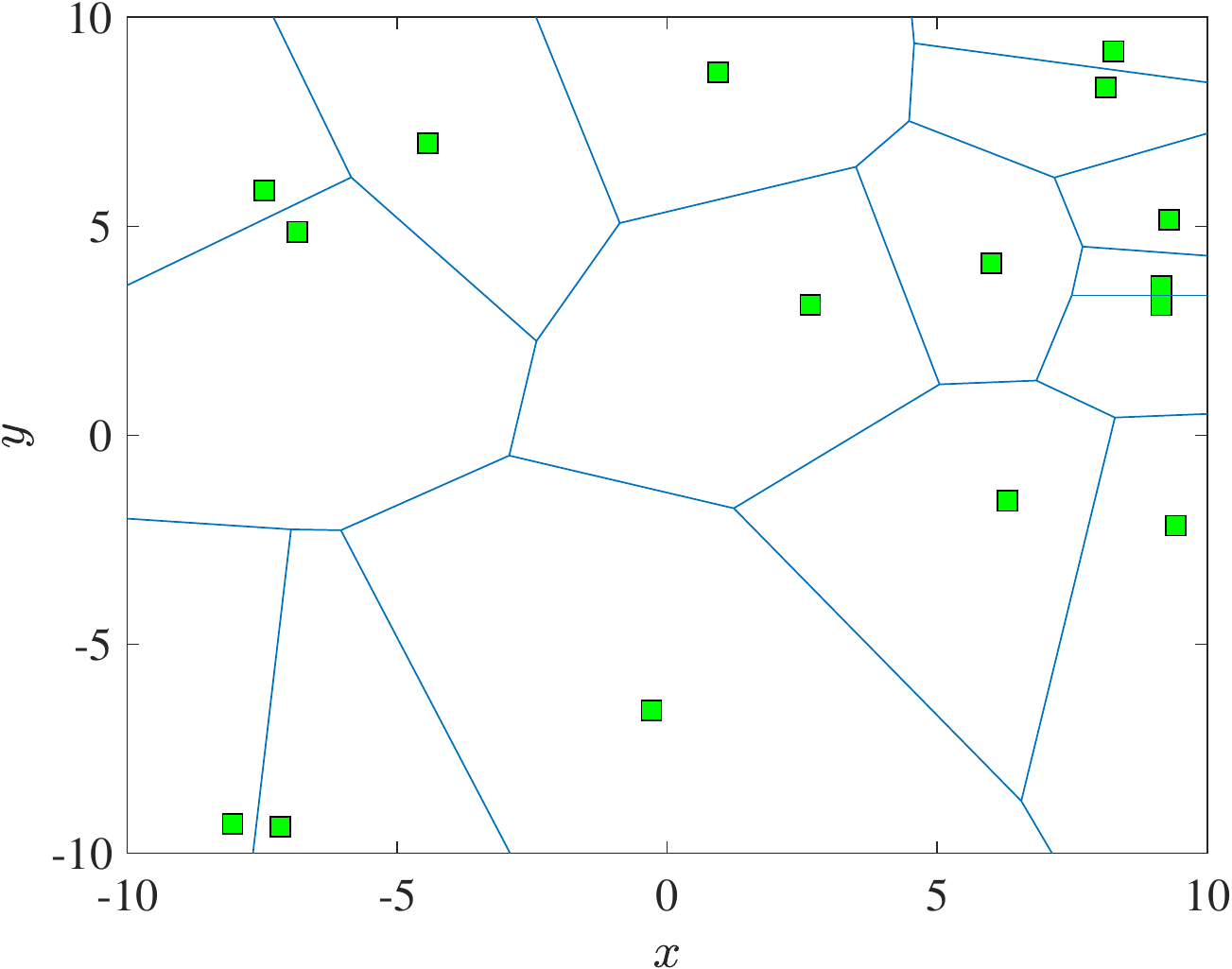}
						\caption{PPP}
				\label{PPP}
		\end{subfigure}%
				\caption{Different types of 2D cell deployments.}
				\label{cellDep}
\end{figure*}

One of the key roles of cellular networks is to address the issue of spectrum congestion. Cellular networks can be formed through different types of cell deployments. 
The mainstream cell deployments can be categorized according to their dimensions. A typical one-dimensional (1D) cell deployment is the Wyner model, which arranges APs in a linear array \cite{5951699}. This model is mostly used for outdoor scenarios along a street, highway or railway \cite{7899477}. Three-dimensional (3D) deployments are studied in some special scenarios such as drone systems \cite{8647225}. For an indoor scenario, which is the main focus of this paper, two-dimensional (2D) grid-based cell deployments are preferable as they can provide a general form of coverage. LED lamps, which act as LiFi APs, are normally fixed on the ceiling of a room. Three types of 2D cell deployments are commonly seen in literature: lattice, hexagon and Poisson point process (PPP) \cite{7362097}. Fig. \ref{cellDep} presents these deployments, which are introduced and discussed below.

\subsubsection{Hexagon}
The hexagonal deployment, which is shown in Fig.~\ref{Hex}, is widely used as an ideal structure of cellular networks, due to the ready implementation of frequency reuse. This deployment provides the highest SINR coverage probability in LiFi \cite{7362097}. Although it is unlikely to have such a deployment in a realistic scenario, studying this deployment is important for providing an upper bound analysis. 

\subsubsection{Lattice}
The lattice deployment is the most practical and commonly used for LiFi, due to the topological nature of ceiling lamps. This deployment, which is shown in Fig. \ref{square}, has been widely adopted in many studies related to HLWNets \cite{7876858,8123892,7274270}. The research in \cite{7362097} proves that the signal-to-interference-plus-noise ratio (SINR) coverage probability of the lattice deployment is very close to that of the hexagonal deployment.

\subsubsection{Poisson Point Process}
Alternative to regular and deterministic deployments, the PPP deployment mimics randomly located APs, as shown in Fig.~\ref{PPP}. It is uncommon and impractical to use this deployment for LiFi, due to the difficulty in fulfilling illumination requirements. Apart from that, PPP offers the worst SINR coverage probability among the noted three deployments \cite{7362097}. Therefore, only a few studies, e.g. \cite{7921566}, use the PPP deployment for LiFi networks.

\subsection{Multiple Access} \label{sec:MA}
The well-known standards for WiFi are IEEE 802.11 series, which adopt carrier-sense multiple access with collision avoidance (CSMA/CA). The standardization of LiFi is still in progress, mainly in three task groups: IEEE P802.15.13, ITU G.vlc and IEEE 802.11bb \cite{ieee802bb}. In this paper, we introduce several multiple access schemes that are potential for LiFi, including CSMA/CA \cite{CSMA_HWL}, time-division multiple access (TDMA) \cite{7343497,7510823}, orthogonal frequency-division multiple access (OFDMA) \cite{8279493,7247378,7820066}, space-division multiple access (SDMA) \cite{7857700,7249135}, and non-orthogonal multiple access (NOMA) \cite{8647397}.

\subsubsection{CSMA/CA}
According to this scheme, mobile terminals or nodes try to avoid collision by sensing the channel and then transmitting the data packet if the channel is found to be idle. CSMA/CA can optionally benefit from the request-to-send (RTS) and clear-to-send (CTS) exchange between the transmitter and receiver \cite{IEEEMACStand}. Once a node is allowed to access the channel, it can use the whole bandwidth.  

\begin{figure*}[t]
	\centering
	\resizebox{0.95\linewidth}{!}{
		\includegraphics{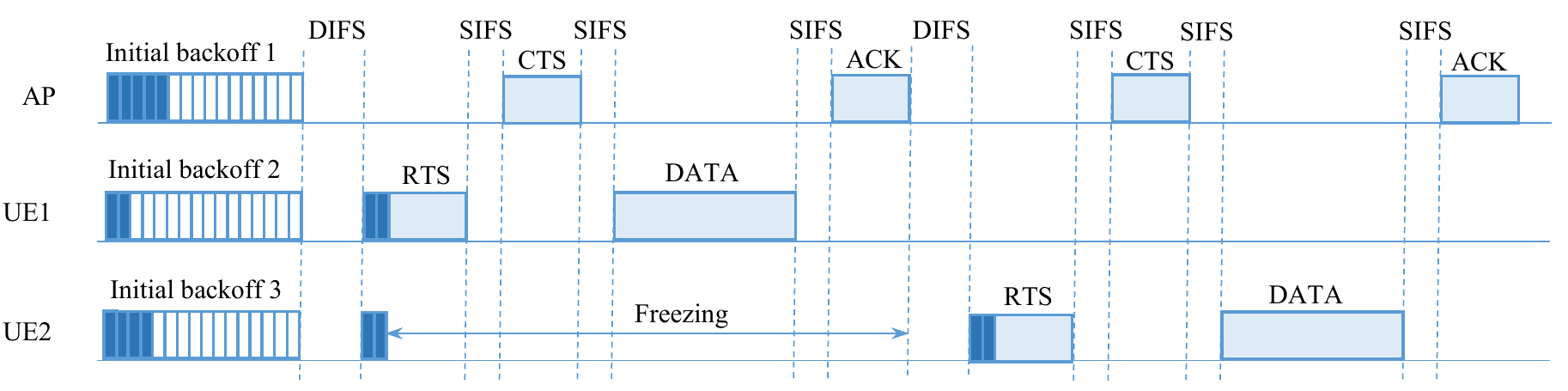}}
	\caption{CSMA/CA with four-way handshaking RTS/CTS mechanism.} 
	\label{figuCSMA}  
\end{figure*}

CSMA/CA adopts a slotted binary exponential backoff scheme to reduce collisions due to the simultaneous transmission of nodes. This is known as the collision avoidance feature of the protocol. Hence, prior to transmission, nodes listen to the channel for a time interval called distributed inter-frame space (DIFS). Then, if the channel is found to be idle, the nodes generate random backoffs, $\mathcal{B}_j$, for $j=1,2,\ldots,N$, where $N$ is the number of competing nodes being served by the associated AP. The value of $\mathcal{B}_j$ is uniformly selected in the range of $[0, w-1]$, where $w$ is the current contention window size. Before the first transmission, $w$ is set to the minimum contention window, $W_{\rm{min}}$, and after each unsuccessful transmission, $w$ is doubled up to the maximum contention window, $W_{\rm{max}}$. 
The node with the lowest backoff is prior to transmit and it sends the RTS frame to the AP before other nodes. If the RTS frame is received at the AP successfully, the AP replies after a short inter-frame space (SIFS) with the CTS frame. Once the node receives the CTS packet, it will proceed to transmit the data frame after the time interval defined by SIFS. Eventually, an acknowledgment (ACK) packet is transmitted after another SIFS seconds by the serving AP to notify successful packet reception. 
Fig. \ref{figuCSMA} illustrates the CSMA/CA with the four-way handshaking RTS/CTS mechanism. CSMA/CA can be used in both WiFi and LiFi systems of HLWNets. However, the research work in \cite{8302445} finds that the conventional CSMA/CA would cause overwhelming collisions if it is used for LiFi in a HLWNet. This is due to the use of different transmission medium in downlink and uplink of LiFi, visible light and infrared, respectively. Hence, a modified CSMA/CA is proposed in \cite{8302445} to reduce the number of collisions, which can be applied to the LiFi system of HLWNets. In the modified version, a channel busy tone is transmitted from the AP to the users to indicate that the channel is already occupied. By means of the modified CSMA, the probability of collision can be reduce to less than $1\%$ \cite{MDS2019Thesis}.

\subsubsection{TDMA}
\begin{figure}[b!]
	\centering
		\begin{subfigure}[b]{0.5\columnwidth}
			\centering
			\includegraphics[width=1\columnwidth,draft=false]{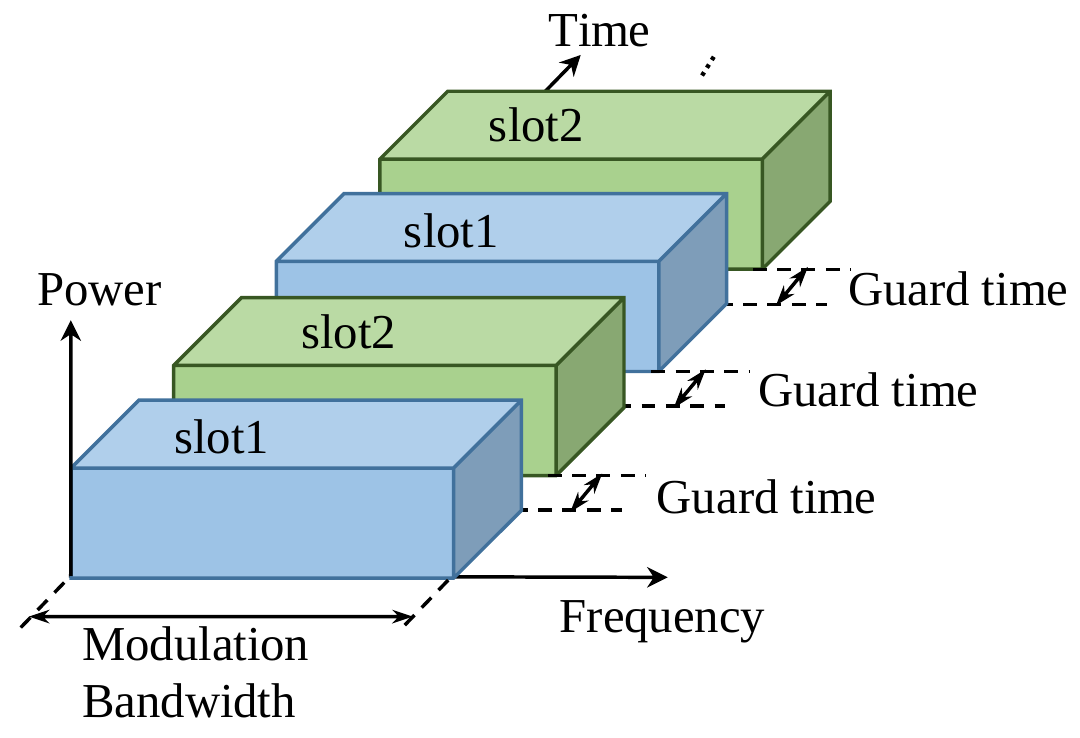}
						\caption{TDMA}
				\label{sub1:TDMA}
		\end{subfigure}%
 \!\!\!\!
		\begin{subfigure}[b]{0.5\columnwidth}
			\centering
			\includegraphics[width=1\columnwidth,draft=false]{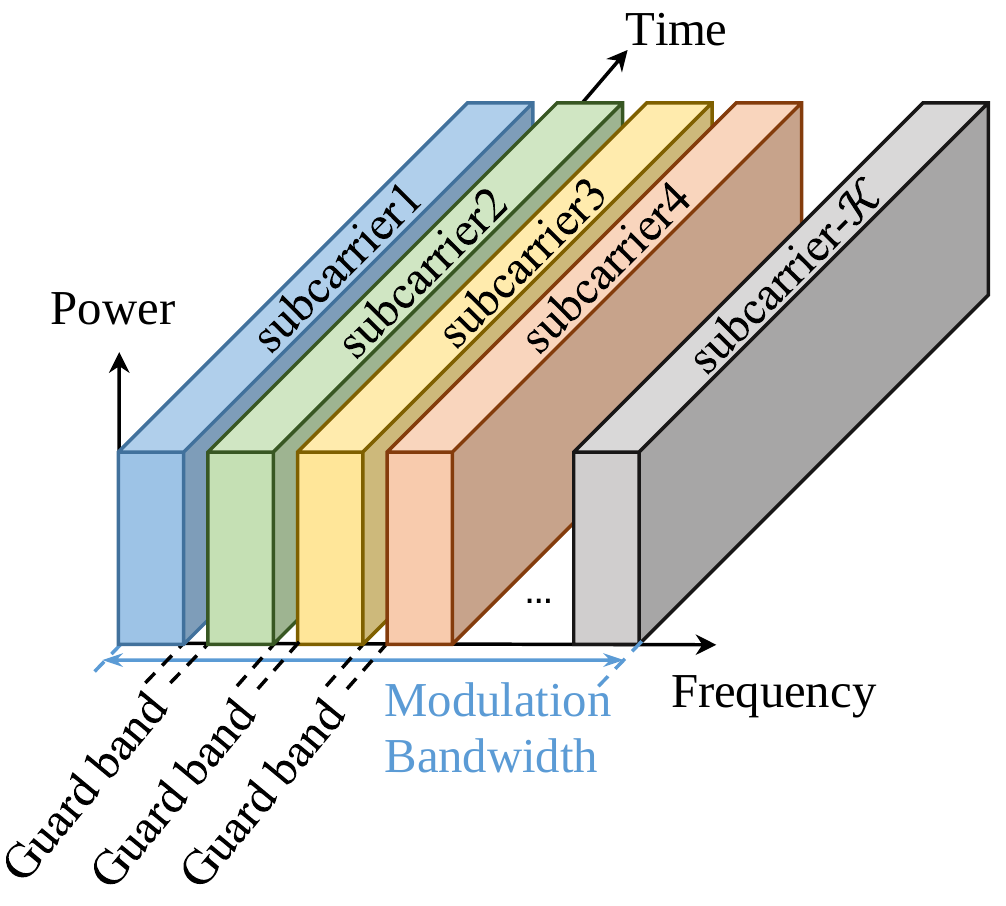}
						\caption{OFDMA}
				\label{sub2:OFDMA}
		\end{subfigure}
	\\ 
	\caption{A comparison between TDMA and OFDMA. }
	\label{FigMA}
\end{figure}

TDMA allows orthogonal access to the whole available modulation bandwidth to all users by assigning various time slots to each of them. Therefore, the users transmit data in rapid succession, one after the other while each uses its own assigned time slot. Fig.~\ref{sub1:TDMA} shows the utilization of bandwidth with TDMA. 

This multiple access technique can be directly used in a HLWNet, particularly in downlink. However, TDMA requires synchronization and it is more difficult for uplink, especially for mobile UEs, due to different signal propagation times of randomly located terminals. These random delays must be compensated by schemes such as timing advance to synchronize the transmission. 
Moreover, the performance of TDMA can be significantly degraded in HLWNets due to inter-channel interference (ICI), that is reported in \cite{7360112}. Therefore, interference mitigation techniques are required to alleviate the ICI which degrades the cell-edge users' performance. TDMA is the most popular multiple access mechanism that is used in the literature, see \cite{7274270,7437374,7056535,7876858} and references therein. 

\subsubsection{OFDMA}
This multiple access mechanism enables users to use frequency resources at different subcarriers and it has been widely considered and implemented in the downlink of long-term evolution (LTE) systems. Fig.~\ref{sub2:OFDMA} illustrates the concept of subcarrier utilization by the OFDMA technique. Multiple access can be achieved in OFDMA by allocating different subsets of subcarriers to individual users. Note that based on the required quality of service (QoS), the achievable data rate can be controlled individually for each user. This can be realized through assigning various numbers of subcarriers to different users. Therefore, OFDMA has been recently considered as a promising and practical option for downlink transmission. OFDMA can be used as the downlink multiple access mechanism of both WiFi, for example WiFi-6 which employs IEEE 802.11ax \cite{7422404}, and LiFi \cite{7876858} in a HLWN. OFDMA has been applied in few HLWNs studies \cite{8357657,7921566} to allocate frequency subcarriers or chunk between users.  

\subsubsection{NOMA} \label{sec:NOMA}
NOMA, also known as power domain multiple access, has been recently gained attention as a promising candidate for $5$G wireless cellular networks, because of its increased spectrum efficiency as well as higher per individual data rate, compared to orthogonal multiple access techniques \cite{7973146}. In contrast to orthogonal multiple access such as TDMA and OFDMA, where users are assigned exclusive time or frequency resources, in NOMA, all users share the same time and frequency resources. 
In fact, users are multiplexed in the power domain using superposition coding and by means of successive interference cancellation (SIC) at the receiver, transmitted signals are decoded. Hence, channel state information (CSI) is needed. In NOMA, users who have good channel conditions are allocated lower transmit power while those with poor channel condition are assigned higher transmit power. HLWNs with NOMA in both LiFi and WiFi systems are studied in \cite{8641243,8647397}, where user clustering is investigated through the coalitional game theory and a merge-and-split algorithm to
determine the optimal user clustering is presented in \cite{8641243}. In \cite{8647397}, a cooperative NOMA based LiFi/WiFi with simultaneous wireless information and power transfer is proposed. Near User receive information and harvests the energy from the LED. Then, the harvested energy is utilized 
to forward the information through WiFi to users who cannot directly communicate with the LED.




\subsection{Modulation Techniques for LiFi}
Typical modulation techniques that suit LiFi systems can be classified into two categories: single carrier and multiple carriers. On-off keying (OOK), pulse position modulation, unipolar pulse amplitude modulation (PAM), pulse width modulation and pulse amplitude modulated discrete multi-tone modulation are common single carrier modulation techniques. These single carrier modulation methods are facile to implement and are ideal for low-speed applications.
Among them, OOK modulation is more popular for low-medium data rate transmission due to its low complexity. By means of OOK, the a bit ``1" is denoted by an optical pulse while a bit ``0" is denoted by the absence of an optical pulse. The return-to-zero (RZ) and non-return-to-zero (NRZ) schemes are the common forms of OOK. In the NRZ approach, the bit duration is the same as the duration of the transmitted pulse to represent ``1" while in the RZ scheme, they are not equal, i.e., the pulse occupies only a partial duration of the bit ``1".
The low complexity implementation of OOK has led to its utilization in commercial optical wireless systems
such as fast IR links operating below $4$ Mbit/s \cite{IRstandard}. 

\begin{figure}[t!]
	\centering
	\resizebox{1\linewidth}{!}{
		\includegraphics{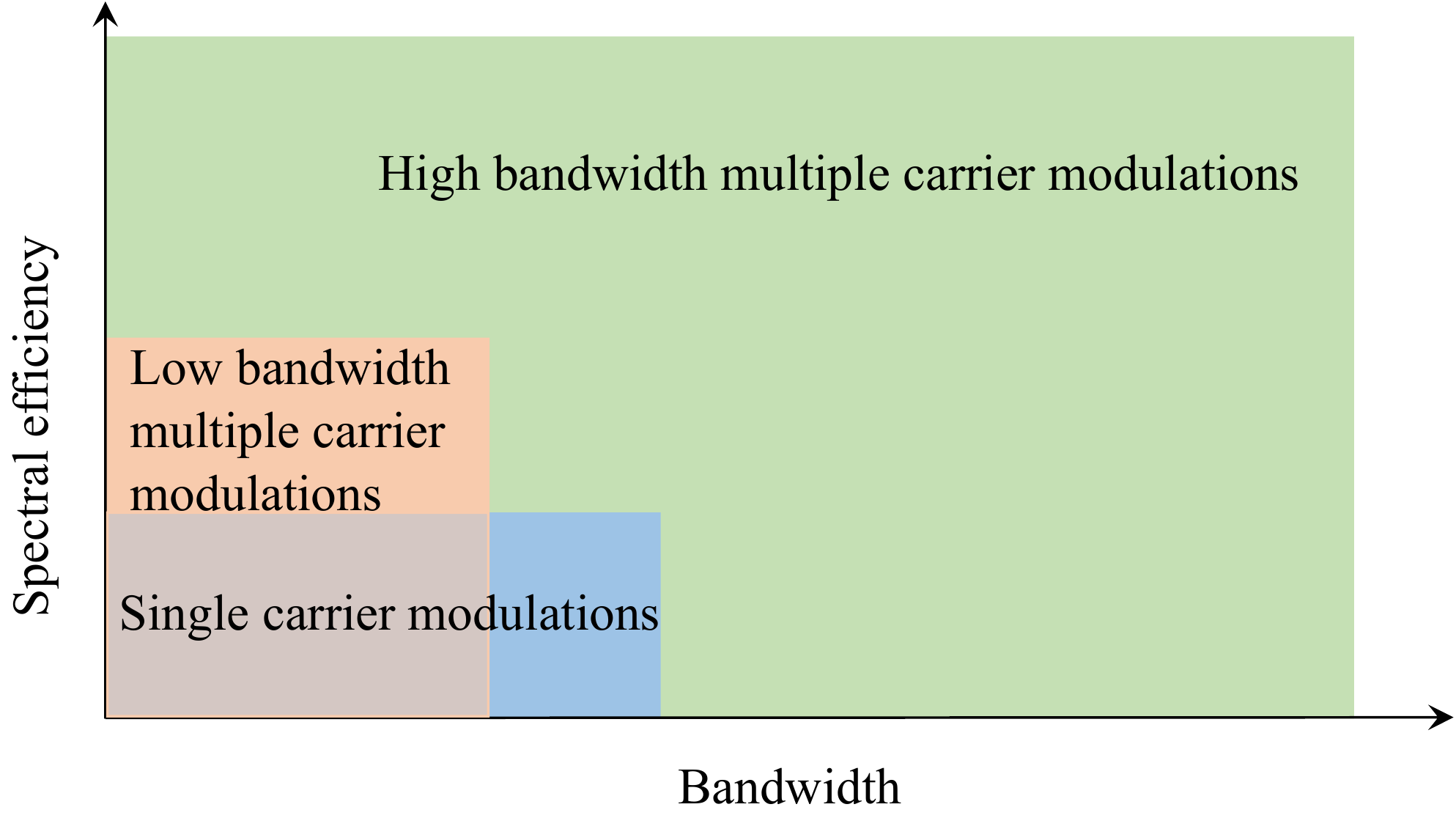}}
	\caption{A comparison of spectral efficiency between single carrier and multiple carrier modulations.} 
	\label{figSE}
	\vspace{-0.1cm}  
\end{figure}

It is noted that single carrier modulation schemes are prone to inter-symbol interference (ISI) \cite{7360112}. Hence, as the required data rate increases, implementing these modulation techniques necessities more complex equalizers at the receiver in order to combat the ISI over the dispersive channel.
In comparison with single carrier modulation schemes, multi carrier modulation techniques are more spectral efficient and can offer higher data rates. Fig.~\ref{figSE} compares single carrier and multiple carrier modulations in terms of spectral efficiency. 
Orthogonal frequency division multiplexing (OFDM) is one of the most common and widely used multiple carrier modulation methods. OFDM is one effective solution to combat the effect of ISI in LiFi systems \cite{1649137}. In LiFi systems, the introduced ISI is a result of passing the signal through a dispersive optical channel at high-data-rate transmission and also using off-the-shelf bandwidth limited LEDs \cite{6249713,6701327}. Benefits of using OFDM include: 1) efficient use of spectrum, 2) robustness against frequency selectivity of the channel by splitting it into narrowband flat fading subcarriers, 3) simple channel equalization by using a single-tap equalizer (while adaptive equalization techniques are being used in single carrier modulation schemes), and 4) it is computationally efficient by using fast Fourier transform (FFT) and inverse FFT (IFFT) techniques. 

In LiFi systems, the modulated signal is used to derive the LED hence it should be positive and real value. Therefore, the conventional OFDM modulator, which generates bipolar and complex signals, cannot fit the intensity modulation and direct detection (IM/DD) requirements \cite{6399084}.
Optical-OFDM (O-OFDM) is a unipolar solution that can be adopted in IM/DD-based transmission. There are several types of O-OFDM that can generate real and non-negative signals. 
Two well-known types of O-OFDM are: direct current (DC)-biased optical OFDM (DCO-OFDM) \cite{490239,1499615} and asymmetrically clipped optical OFDM (ACO-OFDM) \cite{armstrong2006power}.

\begin{figure*}[t!]
	\centering
	\resizebox{1\linewidth}{!}{
		\includegraphics{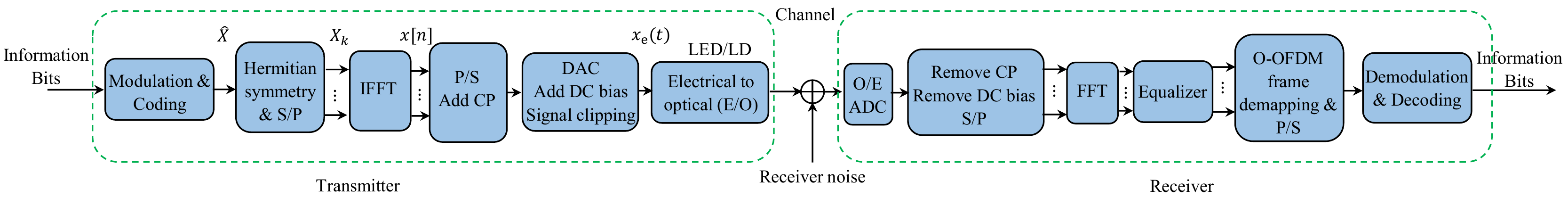}}
	\caption{Illustration of DCO-OFDM system.} 
	\label{figOFDM}
\end{figure*}

\subsubsection{DCO-OFDM}
Fig.~\ref{figOFDM} shows the block diagram of a DCO-OFDM system and its key elements. In a DCO-OFDM system, first the information bits are mapped to quadrature amplitude modulation (QAM) symbols. Let's denote the total number of subcarriers as $\mathcal{K}$. Then, each $(\frac{\mathcal{K}}{2}-1)$ consecutive modulated symbols are grouped to form an OFDM frame to be used as the input of the IFFT module. An OFDM frame can be expressed as:
\begin{equation}
\boldsymbol{X}=[X_0, X_1, ..., X_{\mathcal{K}-1}],
\end{equation} 
where $X_k$ for $k=0,\cdots,\mathcal{K}-1$ are modulated data symbols transmitted on $k$th OFDM subcarrier. To generate real value signals in the time domain, Hermitian symmetry is applied to the OFDM frame, which specifies the following conditions:
\begin{subequations}
\begin{equation}
\label{Hermitian}
X_{\mathcal{K}-k}=X_k^*  \ \ \ \ {\rm{for}}\ \ \  0<k<\frac{\mathcal{K}}{2},
\end{equation}
\begin{equation}
X_0=X_{\mathcal{K}/2}=0,
\end{equation}
\end{subequations}
where $(\cdot)^*$ denotes the complex conjugate operator. After the IFFT operation, time-domain samples are given as:
\begin{equation}
x[n]=\frac{1}{\sqrt{\mathcal{K}}}\sum_{k=0}^{\mathcal{K}-1}X_k{\rm{exp}}\left(\frac{j2\pi kn}{\mathcal{K}} \right), \ \ \ \ \ \ 0\leq n\leq \mathcal{K}-1.
\end{equation}
After passing through the IFFT module, a cyclic prefix (CP) will be added to the samples in order to combat the ISI due to the dispersive wireless channel. After adding the CP, the samples will be fed into a digital to analog converter (DAC) module. A DC bias will be added to the analog waveform to ensure the modulated signal, $\tilde{x}(t)$, must be positive. The positive constraint is required for optical systems that perform intensity modulation. Therefore, 
\begin{equation}
x_{\rm{e}}(t)=x_{\rm{DC}}+\tilde{x}(t), 
\end{equation}
where 
\begin{equation}
x_{\rm{DC}}=\eta\sqrt{\mathbb{E}[\tilde{x}^2(t)]}, 
\end{equation}
and $\eta$ is the conversion factor. In general, the condition $\eta=3$ guarantees that less than $1\%$ of the signal is clipped. In this case, the clipping noise is negligible \cite{6399084}. The current signal $x_{\rm{e}}(t)$ derives the LED to generate the optical signal $x(t)$. 

\begin{table*}[t!]
\renewcommand{\arraystretch}{1.2}
\caption{A comparison between ACO- and DCO-OFDM techniques. }
\label{table:ACO-DCO-Comparison}
\centering
\begin{tabular}{|l|l|l|l|l|l|}
\hline
 & \vtop{\hbox{\strut Spectrum}\hbox{\strut efficiency}} & \vtop{\hbox{\strut Energy}\hbox{\strut efficiency}} & \vtop{\hbox{\strut Data}\hbox{\strut rate}} & \vtop{\hbox{\strut Hardware}\hbox{\strut cost}} & \vtop{\hbox{\strut Robust against}\hbox{\strut multipath}}\\
\hline
\multirow{1}{*}{\vtop{\hbox{DCO-OFDM}}} & Medium & Low & High & Medium & High   \\
\hline
\multirow{1}{*}{\vtop{\hbox{ ACO-OFDM}}} & Low & High & Medium & Medium & High  \\
\hline
\end{tabular}
\end{table*}

\subsubsection{ACO-OFDM} 
ACO-OFDM is another type of energy-efficient O-OFDM that can prevent adding a DC bias to the signal.  
In ACO-OFDM, only odd subcarriers are used to bear information which results in a loss of spectral efficiency. Hence, the OFDM frame is $\boldsymbol{X}=[0, X_1, 0, ..., X_{\mathcal{K}-1}]$. Furthermore, the elements of $\boldsymbol{X}$ should fulfill the Hermitian symmetry defined in \eqref{Hermitian}. Compared to DCO-OFDM, half of the spectrum is sacrificed by ACO-OFDM to make the time-domain signal unipolar. Therefore, the signal generated after IFFT is an anti-symmetric real value given as:
\begin{equation}
x[n+\mathcal{K}/2]=-x[n], \ \ \ \ \ \ 0< n<\mathcal{K}/2 .
\end{equation} 
The anti-symmetry property of $x[n]$ guarantees that no information data is lost due to signal clipping at the zero level. A detailed comparison of ACO- and DCO-OFDM techniques is provided in \cite{6415964}. It is shown that the BER of the ACO-OFDM scheme is half of the DCO-OFDM one for a given constellation due to transmitting data symbols only on half of the subcarriers. 
A comparison between ACO- and DCO-OFDM in terms of spectrum/energy efficiency, data rate, hardware complexity and robustness against multipath is provided in Table~\ref{table:ACO-DCO-Comparison}.

\subsection{Illumination Requirements}
\begin{figure}[t!]
	\centering
	\resizebox{1\linewidth}{!}{
		\includegraphics{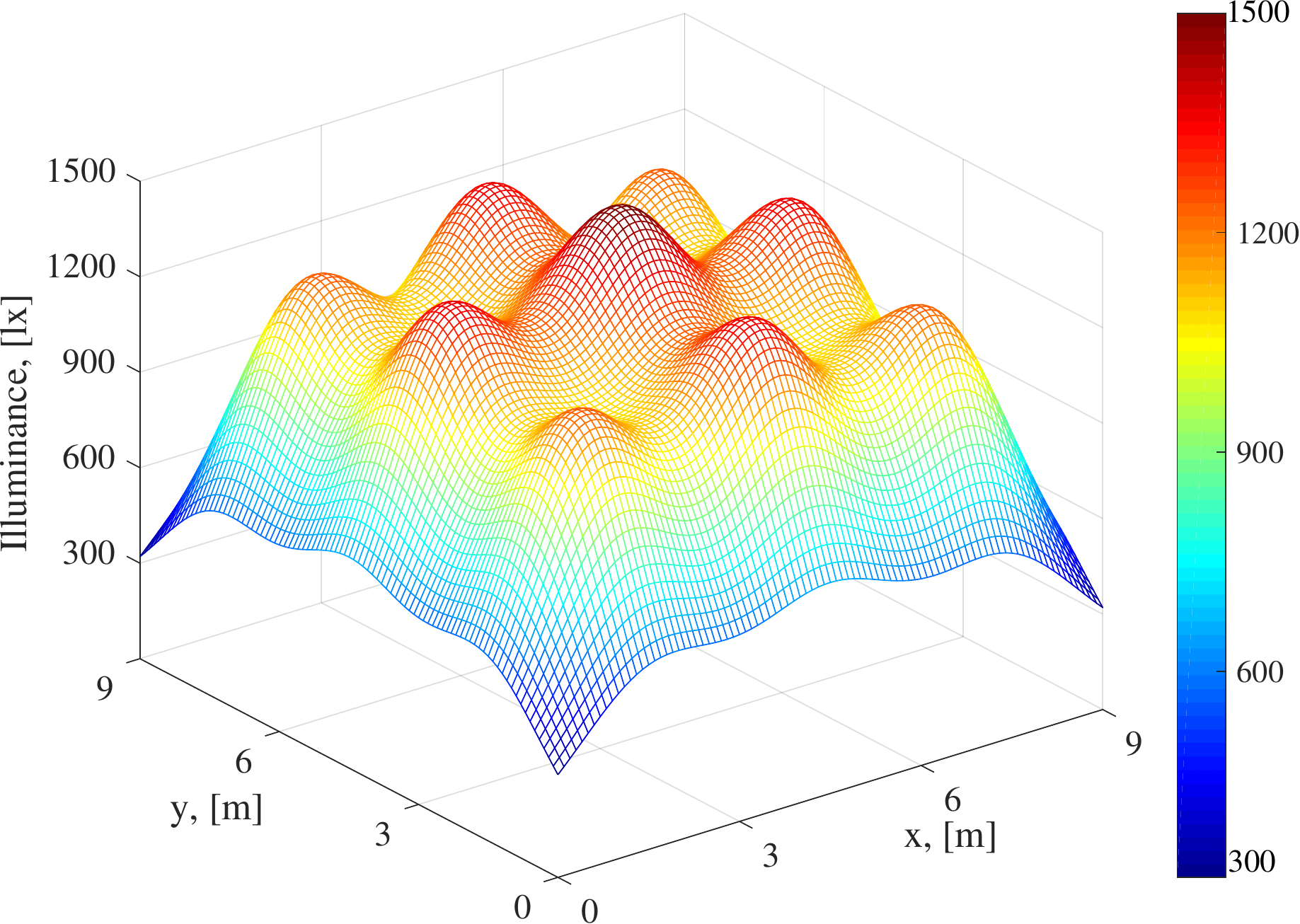}}
	\caption{The distribution of illumination at different positions of the room.} 
	\label{figillumination}
\end{figure}

One effective way of improving energy efficiency of HWLNets is a proper design of the placement or layout of the LEDs within the considered indoor LiFi system. 
By optimizing the locations and controlling output power levels of the LEDs, the minimum energy consumption and at the same time a desired illumination pattern can be achieved.  
It is very important that, when designing an energy efficient HLWNet, the illumination constrain of the room should also be satisfied. According to the international organization for standardization (ISO) on light and lighting, the illumination requirements for an indoor environment should be in the range of $300$ to $1500$ lx \cite{1277847}. Typically, the illuminance is low at the corners of a room while it is high at the center of the room. Hence, a careful design is required to provide uniform illuminance to meet the ISO requirement. According to \cite{1277847}, the illuminance level at each location can be  obtained as follows:
\begin{equation}
    I_{\rm Lum}(x,y)=\sum_{i=1}^{N_{\rm LED}}\frac{I_0}{d_i^2}\cos^m(\phi_i)\cos(\psi_i),
\end{equation}
where $I_0$ is the center luminous intensity of LEDs, which is assumed to be the same for all LEDs, $d_i$ is the Euclidean distance between $i$th LED and the arbitrary location $(x,y)$; the number of LEDs is denoted by $N_{\rm LED}$ and the Lambertian order, $m$, can be obtained based on the half-intensity angle, $\Phi_{1/2}$ as $m=-1/\log_2(\cos(\Phi_{1/2}))$. The radiance and incidence angles at the location $(x,y)$ are denoted by $\phi_i$ and $\psi_i$. 
Fig.~\ref{figillumination} shows the distribution of illumination at different positions in a room of size $9\times9$ m$^2$ with $9$ APs. As we expected, the illuminance level is minimum at the corners, however, the separation between LEDs are designed in a way that the minimum illumination requirement of $300$ lx is satisfied. For this simulation, we assume $I_0=1.2$ cd, the vertical distance of LEDs and $xy$-plane is $2.15$ m and the separation between LEDs is $3$ m. 

Simultaneous energy efficient design and illumination constraints have been considered in several works \cite{EEandIllum2016,EEandIllum2018}. Specifically, in \cite{EEandIllum2016}, the authors studied the problem of designing an energy efficient LED layout with the consideration of illumination constraints. Hence, a constrained optimization problem is formulated, which achieves minimum power consumption while providing a nearly uniform illumination throughout the indoor environment. In \cite{EEandIllum2018}, a fast game-theory-based algorithm is formulated to maximize the energy efficiency under the  illumination constraints. The generalized Nash equilibrium model is chosen to analyze the competition among VLC pairs to achieve the maximum energy efficiency. Therefore, it is important to consider the illumination in the design of HLWNets not only to fulfill the illumination requirement of the room but also to provide an energy efficient hybrid network.

\subsection{Backhauling}
In cellular networks, the links that connect the APs to the central unit are called backhaul links. With the rapid growth of mobile users, the design and implementation of cost efficient backhaul structures have become a crucial challenge in cellular networks \cite{5185525}. Furthermore, backhaul networks have had to grow accordingly.  
Radio interface has been considered as the bottleneck of cellular networks for a long period of time due to their low efficiency. However, this bottleneck has been shifted to backhaul networks in recent years along with the improved radio interface and demand for high data-rate service and throughput. 

To design and implement a cost-efficient backhaul in a HLWNet is a possible challenge, where separate or common backhauls for LiFi and WiFi networks that can guarantee availability, scalability and flexibility of HLWNets should be considered. 
LiFi networks consist of multiple small cells known as attocells \cite{haas2013high}. With recent and modern interference management techniques, the spectrum can be fully reused at each LiFi attocell \cite{8387504}. Therefore, the LiFi APs are able to provide several Gbps data rate \cite{SPIE_2017}. Hence, the backhaul links should be able to transfer hundreds of Gbps in LiFi networks. Various technologies such as fixed copper, fiber-optical links over passive optical network \cite{manis2005passive} and optical wireless technologies such as visible light \cite{7996637} or infrared \cite{8558716} are studied in the literature as possible solutions to provide backhaul connectivity. The studied methods in the literature are listed in Table~\ref{table:backhaul_approaches}.

\begin{table*}[t]
\renewcommand{\arraystretch}{1.2}
\caption{A summary of various approaches for backhaul in HLWNets.}
\label{table:backhaul_approaches}
\centering
\begin{tabular}{|l|l|l|l|l|l|}
\hline
 & foundation & Data rate & cost & Flexibility & Ref. \\
\hline
\multirow{3}{*}{\vtop{\hbox{Wired}}} & \multirow{1}{*}{Copper/PLC} & Medium & Medium & Low & \cite{1205458,1613633,7052407,6524532} \\
\cline{2-6}
 & \multirow{1}{*}{Ethernet/POE} & Medium & Medium & Low  & \cite{mark2014ethernet} \\
\cline{2-6}
& \multirow{1}{*}{Fiber} & High & High & Low  & \cite{7193335} \\
\hline
\multirow{3}{*}{\vtop{\hbox{Wireless}}} & \multirow{1}{*}{mmWave} & High & Medium & High & \cite{7289244}  \\
\cline{2-6}
 & \multirow{1}{*}{Infrared} & High & Medium & High & \cite{8558716} \\
\cline{2-6}
&\multirow{1}{*}{Visible light} & High & Medium & High & \cite{7996637,8558716}  \\
\hline
\end{tabular}
\end{table*}

The idea of using the existing wiring infrastructure within buildings for backhaul, known as power-line communication (PLC), was initially proposed by Komine and Nakagawa in \cite{1205458}. Later works on hybrid PLC-VLC include: performance evaluation of narrowband OFDM on an integrated system of PLC and VLC \cite{1613633}, broadband communication over PLC-VLC system for indoor applications \cite{7052407} and LiFi integrated to PLC for smart illumination and communication purposes \cite{6524532}. Ethernet is another option for wired backhaul based on the power-over-ethernet (POE) standard. A dual-hop relaying transmission over Ethernet and VLC is proposed in \cite{mark2014ethernet}, which is a cascaded system of POE and VLC. 
Another option for wired backhaul is optical fiber. The huge bandwidth of fiber permits for multi-Gbps transmission, however, it is expensive and time consuming to lay down optical fiber to all households \cite{7193335}. 

The wireless solutions of the backhaul are discussed in \cite{7996637,8558716,8647686,7289244}. Millimeter wave is proposed as a possible solution for point-to-point backhaul connectivity in \cite{7289244}, which can support multi-Gbps and it is cost efficient compared to fiber optic solution. Optical wireless communication is another possible solution, which is capable of handling future high-speed backhaul networks. 
In \cite{7996637}, a wireless backhaul solution based on visible light communication is proposed for indoor LiFi attocell networks.
The authors consider full frequency reuse and in-band techniques for bandwidth allocation between the access and backhaul links. Furthermore, relaying protocols such as amplify-and-forward and decode-and-forward are studied to realize dual-hop transmission. 
In \cite{8558716}, both visible light and infrared bands are evaluated as the backhaul solutions of LiFi networks. The effects of inter-backhaul and backhaul-to-access network interference on the SINR  are analyzed. To realize backhauling with visible light, the authors assessed both full-reuse and in-band bandwidth allocation techniques. 
A HLWNet in which LiFi and WiFi systems are served by the same backhaul link is investigated in \cite{8377176}. Resource allocation problems are studies to enhance the users' fairness index by optimizing the power allocation between LiFi and WiFi systems in the HLWNet with the common backhaul.

\section{Performance Metrics} \label{sec:performance_metrics}
In order to evaluate the performance of HLWNets several metrics including coverage probability, spectral efficiency, area spectral efficiency, network throughput, fairness and quality of service are considered in various literature. In the following, this metrics and the related studies are discussed in detail. 
\subsection{Coverage Probability}
The coverage probability is defined as the probability that SINR is higher than a certain threshold in a cellular network. In HLWNets, particularly in the LiFi systems, the coverage probability depends not only on the location of a UE but also on its orientation. The orientation can be determined uniquely by three angles, $(\alpha,\beta,\gamma)$ (It will be discussed in more detail in Section~\ref{Sec:DeviceOrientation}). Therefore, the coverage probability of a HLWNet for a given device orientation can be determined mathematically as follows \cite{6042301}:
\begin{equation}
    P_{\rm c}=\Pr\{\mathcal{S}>\mathcal{S}_{\rm t}\left|{(\alpha,\beta,\gamma)}\right.\},
\end{equation}
where $\mathcal{S}$ is the received SINR at the UE and $\mathcal{S}_{\rm t}$ is the considered SINR threshold. Hence, the coverage probability for a given device orientation can be interpreted as:
i) the probability that the SINR of a randomly-chosen user is higher than the threshold, $\mathcal{S}_{\rm t}$,
ii) the average number of users whose SINR is greater than the threshold, $\mathcal{S}_{\rm t}$ at any time over the total number of users,
iii) the fraction of the cellular area where the SINR is greater than the threshold, $\mathcal{S}_{\rm t}$ \cite{7925676}.
It can be deduced that the probability of coverage is equal to the complementary cumulative distribution function of SINR.  

Coverage probability has been analyzed in several hybrid LiFi and WiFi studies including  \cite{7145863,8279493}.
Specifically, the authors in \cite{7145863} considered a HLWNet where the coverage area has been improved by means of WiFi APs and the hybrid system is able to provide ubiquitous coverage. Coverage and rate analyses, particularly for downlink of hybrid WiFi and LiFi cellular networks, are studied in \cite{8279493}. The authors provide a stochastic geometry framework to analyze the coverage area, where the framework can be configured to assess the coverage performance in a standalone RF network, VLC-only system, opportunistic RF/VLC network as well as the hybrid one. 

\subsection{Fairness}
The fairness criterion is a measure to evaluate how fairly the available resources are allocated among the users. In fact, when dealing with resource allocation, this metric becomes of interest to assess the fair assignment of available resources. There are several fairness definitions such as Jain's fairness index \cite{jain1984fair}, max-min fairness \cite{5339590}, quality of experience fairness \cite{7588099} and worst case fairness \cite{497885}. Among them,  Jain's fairness index has been widely used by researchers. It can be expressed as follows \cite{jain1984fair}:
\begin{equation}
    I=\frac{\left(\sum_{i=1}^{N}x_i\right)^2}{N\sum_{i=1}^{N}x_i^2},
\end{equation}
where $x_i$ is the average data rate of $i$th user and $N$ denotes the total number of users. Note that the $I$ is a factional value bounded between $1/N$ and $1$. Perfect user fairness corresponds to $1$ while the value $1/N$ corresponds to the case where all available resources are only allocated to one user.  

It is shown that a hybrid RF/LiFi network can remarkably enhance the fairness index among users \cite{7247378,8357657,8013858,7056535}. Specifically, it is expressed in \cite{7247378} that a hybrid LiFi and RF system can improve the fairness index among mobile users by means of a dynamic load balancing algorithm. In \cite{8357657}, the authors proposed a new joint power allocation and load balancing scheme for a hybrid LiFi/RF network consisting of one RF AP and multiple LiFi APs. It is illustrated that the hybrid system is able to enhance the fairness index by transferring the users with poor QoS from overloaded APs to APs with a lower traffic. In \cite{8013858}, the fairness index is enhanced by utilizing a fuzzy logic load balancing algorithm in a hybrid LiFi and WiFi network. The fairness improvement based on their proposed algorithm is greater than the traditional signal strength strategy for AP selection, especially in highly dense networks. The grade of fairness (GoF) for a hybrid LiFi and WiFi network is evaluated in \cite{7056535}. The GoF for LiFi and WiFi systems are defined as follows:
\begin{subequations}
\begin{equation}
    I_{\rm GoF,LiFi}=\left|1-\frac{\epsilon_{\rm T, LiFi}}{\epsilon_{N_{\rm LiFi}}} \right|
\end{equation}
\begin{equation}
    I_{\rm GoF,WiFi}=\left|1-\frac{\epsilon_{\rm T, WiFi}}{\epsilon_{N_{\rm WiFi}}} \right|
\end{equation}
\end{subequations}
where $\epsilon_{\rm T, LiFi}$ and $\epsilon_{\rm T, LiFi}$ are the LiFi and WiFi proportion of total hybrid network throughput, respectively; also $\epsilon_{N_{\rm LiFi}}$ and $\epsilon_{N_{\rm WiFi}}$ are the fraction of LiFi and WiFi connected users, respectively. 
The GoF is usually used to assess the network's average fairness. It is shown in \cite{7056535} that the hybrid system can provide a higher GoF especially when the gap between the LiFi (WiFi) throughput percentage and the fraction of LiFi (WiFi) connected users is low. 

\subsection{Spectral Efficiency}
Spectral efficiency is one of the typical metrics of measuring how efficiently a limited spectrum link is used. Many studies on hybrid RF/VLC networks have repeatedly confirmed that the SE of the hybrid network is much greater than either of RF or VLC standalone networks \cite{7145863,7820066,7247378,7794887}. Specifically, the authors have shown in \cite{7145863} through Monte-Carlo simulations that the hybrid RF/ VLC system can provide higher spectral efficiency and data rate compared to an RF-only network. 
In \cite{7820066}, a two-stage AP selection algorithm for hybrid RF/VLC networks has been introduced based on fuzzy logic techniques, which benefits from a low computational complexity. In \cite{7247378}, a spectral efficient hybrid LiFi and millimeter wave indoor network is proposed to address the dynamic load balancing issue for mobile users. The idea was later extended in \cite{7794887} by jointly optimizing the AP assignment and resource allocation in a hybrid LiFi/RF network, where it is shown that spectral efficiency is improved twice compared to the conventional signal strength strategy. 

\subsection{Area Spectral Efficiency}
Since the AP in a LiFi network covers a confined area known as an attocell, the bandwidth resource can be widely reused in space \cite{7511139}. In order to have a fair comparison between the throughput in a HLWNet and a LiFi or WiFi standalone network, the area spectral efficiency (ASE) is considered in several studies on hybrid LiFi/WiFi networks \cite{7511139,6655152,6965999,7056535}. 
The ASE reflects the performance of the user spectral efficiency for a given user density. Mathematically, ASE can be expressed as:
\begin{equation}
    \zeta=\frac{R}{A_{\rm r}},
\end{equation}
where $R$ is the system data rate and $A_{\rm r}$ is the area of the indoor environment. LiFi networks are capable of providing high ASE compared to standalone RF cellular networks \cite{6655152}. A comparison of ASE between the recent RF systems using femtocell and VLC for indoor communication is presented in \cite{6655152}. It is shown that a LiFi network, due to high bandwidth reuse, is able to achieve several hundred times higher ASE than a femtocell.
Therefore, the integration of both LiFi and RF networks not only can guarantee ubiquitous coverage but also higher ASE.
The authors in \cite{6965999} proposed a hybrid LiFi and femtocell RF system that increases the average ASE by at least two orders of magnitude compared to a standalone RF network. 
In \cite{7056535}, a cooperative load balancing mechanism is proposed, which can achieve higher proportional fairness as well as higher ASE in comparison with a standalone LiFi network. Both distributed and centralized resource allocation schemes are developed and it is shown that a higher ASE without any sacrifice in the fairness can be achieved through the hybrid RF/LiFi system.


\subsection{Energy Efficiency}
With the rapid growth in the request for high data rate communication, the corresponding energy consumption is also increasing quickly. In wireless communication, energy efficiency is defined as the ratio of the transmission data rate and consumed power in bits-per-Joule, that is, \cite{7437374}:
\begin{equation}
    \eta=\frac{R}{P},
\end{equation}
where $R$ and $P$ denote the achievable data rate in [bits/s] and the total transmitted power in [W], respectively. 
Research has shown that hybrid RF/VLC networks are considerably more energy efficient than RF or VLC standalone networks \cite{7437374,8357657,7362622}. 
Specifically, in \cite{7437374}, the energy efficiency of integrating VLC systems with RF networks is studied. The authors formulated an optimization problem of bandwidth and power allocation to maximize the energy efficiency of the hybrid RF/VLC network. The impact of various parameters such as the number of users, line-of-sight (LoS) availability and number of LEDs on the energy efficiency is investigated. The superior performance of a hybrid RF/VLC network compared to an RF-only system is shown. 
In \cite{8357657}, a joint optimization of power allocation and load balancing for hybrid VLC/RF networks is proposed. The authors introduced an iterative algorithm to distribute the transmission power among the serving users. By formulating an optimization problem, their proposed scheme is able to maximize the achievable data rate and consequently guarantee energy efficiency of the hybrid VLC/RF network. It is shown that the new algorithm converges fast and it can provide better performance compared to the conventional subgradient method. In \cite{7362622}, the overall energy efficiency of a mixed RF/VLC network is improved by optimal power allocation. It is shown that the energy efficiency of the hybrid network is $3$ bits-per-Joule greater than an RF standalone system. 

\subsection{Network Throughput}
One of the main aims of HLWNets is to improve users and the network throughput. Almost all studies on the topic of HLWNets have confirmed throughput enhancement compared to a standalone LiFi or WiFi network \cite{7921566,7274270,8279493,7511139,7876858,7056535}. Specifically, in \cite{7921566}, a simple RF deployment is introduced to enhance the per user outage data rate performance\footnote{Outage data rate is defined as the probability that the achieved data rate is less that a threshold value $R_{\rm th}$. Mathematically, it is given as $\Pr\{R\leq R_{\rm th}\}$.} of standalone LiFi systems and improve the overall network throughput.
It is shown that the outage probability can be halved via a hybrid LiFi/RF network compared to a standalone LiFi system for a target throughput of $30$ Mbps. The field of view (FoV) impact on the average user throughput is evaluated in a hybrid LiFi/RF and compared with a standalone LiFi and RF system in \cite{8279493}. It is denoted that the hybrid network can improve the user throughput more than $30$ Mbps and more than $40$ Mbps compared to a standalone LiFi and RF network, respectively. 
In \cite{7056535}, a cooperative load balancing scheme is proposed for hybrid LiFi/WiFi networks. It is shown that the hybrid system is able to enhance the average user throughput for both regular and merged cell\footnote{A merged cell includes several neighbouring cells to support mobility
and reduce the ICI.} formations.  
In \cite{7274270}, the authors proposed a dynamic load balancing mechanism with the consideration of handover for hybrid LiFi and WiFi networks. The utility function of the proposed load balancing algorithm considers both system throughput and fairness and it is shown that the hybrid network can significantly improve the user throughput by serving the mobile users via WiFi APs and quasi-static users via LiFi APs. An analysis of system throughput in the presence of link blockage is carried out in \cite{7511139,7876858}. It is shown that the hybrid RF and LiFi network is able to address the issue of link blockage and enhance the system throughput compared to a standalone LiFi network. 

\subsection{Quality of Service}
The other important metric is the user QoS where it can be leveraged by means of hybrid RF and VLC networks. In fact, QoS is a qualification or measurement of a service performance met by the users in a network. In order to evaluate and quantify the network QoS, several metrics such as bit rate, transmission delay, throughput and link availability can be used. Furthermore, the growing request on delay-sensitive applications, including video streaming and games, necessitates considering other QoS assessments at the data-link layer. Particularly, when the QoS constraints are required, cross-layer evaluation associated to both physical and data-link layers were undertaken by many researchers, see \cite{6587995} and references therein. Hybrid LiFi/WiFi systems are considered as a promising potential to fulfill the future QoS constraints and requirements \cite{8464895,8288604}. 
In \cite{8464895}, the effective capacity is considered, which is a link-layer metric under the QoS constraints such as delay or buffer overflow. Then, the effect of LoS link availability and the number of users on the effective capacity has been evaluated. 
Later, the same authors generalized the idea of assessing system performance under the QoS constraints in \cite{8288604} by evaluating the link selection in a hybrid LiFi/RF system under statistical queueing requirements. By means of numerical results, it is shown that the LiFi system enhances the delay performance of hybrid LiFi/RF networks.


\section{User Behavior Modeling} \label{sec:behavior_modeling}
In a LiFi cellular network, and, as a result in hybrid LiFi and WiFi networks, the user throughput and QoS depend on several factors such as user mobility, device random orientation and link blockage. There are plenty of mobility models proposed in the literature \cite{bai2004survey}. Among them, the random waypoint (RWP) mobility model is one of the most simple and well-known models that is considered for the simulation of user mobility either for indoor or outdoor environments. Device orientation and link blockage are the other two significant factors that can affect the user throughput in LiFi networks. In the following, these factors are discussed. 

\subsection{Mobility Model}

Various indoor mobility models have been considered in \cite{512994,5700140,1381964}. Specifically, the authors in \cite{512994} considered a personal mobility problem in rectangular microcells where users may either move with constant speeds or stay for a call duration. In their mobility model, users tend to move on straight lines until they change their directions according to a Poisson process model. In \cite{5700140}, a novel rule-based mobility model inside a building with several rooms is proposed. The users moves along specific paths from one room to another. 
The constrained mobility model is proposed in \cite{1381964}, where nodes move along the edges of a graph. The edges represent valid paths inside the room. Each node chooses a destination randomly from a subset of graph vertices and moves along the edge towards it. A three-dimensional (3-D) indoor mobility model inside buildings with the consideration of boundary conditions is proposed in \cite{669092}. Similar 3-D mobility for an unbounded indoor building has been introduced in \cite{669091}. In both \cite{669092,669091}, the horizontal (on squared-shaped floors) and vertical movement (at the staircase region) are modeled. The authors extended the same idea to a 3-D mobility model inside high-Rise building environments where vertical motions are done with elevators \cite{892547}. A new and realistic 3-D indoor mobility model according to real building data is introduced in \cite{6547059}. To model a building, various factors such as walls, floors and ceilings are considered. Then, a graph-based model is introduced where users move along the edges between nodes on valid paths. A realistic mobility model for an indoor environment, named the preferred route indoor mobility model, which is based on the hidden Markov model and shortest path algorithm, is introduced in \cite{6784153}. A novel mobility model for indoor conference scenarios is proposed in \cite{6902231}. The authors used a multi-agent event-based simulator based on the crowd behavior to model the movement of users inside the conference room.  Other indoor mobility models can be found in \cite{8116395,bhattacharya2012analytical}. A simple and widely used mobility model which is implemented in software simulators such as NS-3 is the RWP model. Recent studies on HLWNets have considered the RWP mobility model in their simulations \cite{7925839,7820066,7510823}. In the following, this commonly used indoor mobility model is explained. 

\begin{figure}[t!]
	\centering
		\begin{subfigure}[b]{0.65\columnwidth}
			\centering
			\includegraphics[width=1\columnwidth,draft=false]{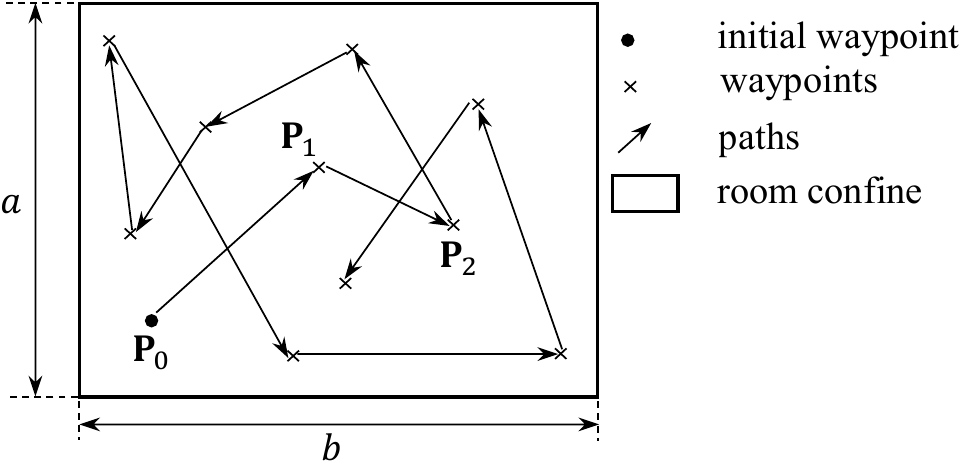}
						\caption{Square room}
				\label{sub1:RWP}
		\end{subfigure}%
 ~
		\begin{subfigure}[b]{0.35\columnwidth}
			\centering
			\includegraphics[width=1\columnwidth,draft=false]{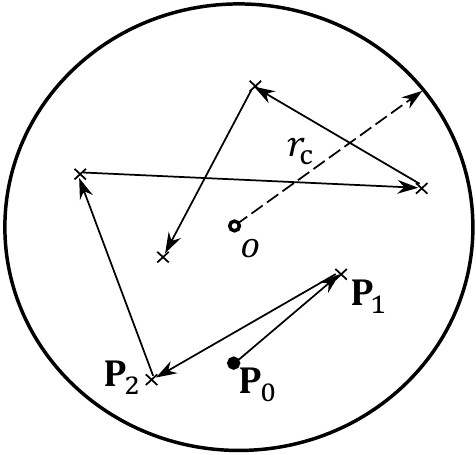}
						\caption{An attocell}
				\label{sub2:RWP}
		\end{subfigure}
	\\ 
	\caption{Random waypoint model in (a): a room of size $a\times b$ and (b) an attocell of radius $r_{\rm c}$. }
	\label{RWP}
\end{figure}

The RWP mobility model was initially introduced in \cite{Broch1998} to model human movement in a random manner. After that, many studies have focused on the RWP to obtain its statistics, see for instance \cite{Bettstetter,hyytia2007random} and references therein. Fig.~\ref{RWP}-(a) shows the basic concept of the RWP mobility model in a room of size $a\times b$.
According to the RWP model, at each waypoint, the UE needs to satisfy a number of properties to move to the next waypoint, these include: i) the random destinations or waypoints that are chosen uniformly with probability $1/ab$; ii) the movement path is a straight line; and iii) the speed is constant during the movement between two consecutive waypoints. Mathematically, the RWP mobility model can be expressed as an infinite sequences of triples: $\{(\textbf{P}_{n-1},\textbf{P}_{n},V_n)\}_{n\in\mathbb{N}}$ where $n$ denotes the $n$th movement period during which the UE moves between the current waypoint $\textbf{P}_{n-1}=(x_{n-1},y_{n-1})$ and the next waypoint $\textbf{P}_{n}=(x_n,y_n)$ with the constant velocity $V_n$. The transition length is defined as the Euclidean distance between two consecutive waypoints as the UE progresses, and is given by $L_n=\|\textbf{P}_{n}-\textbf{P}_{n-1}\|$. Here, the transition lengths $\{L_1, L_2, \ldots\}$ are non-negative independent identically distributed (i.i.d.) random variables.
For a square room of length $a$, the mean of transition lengths is given as $\mathbb{E}[L]=0.5214a$ where $\mathbb{E}[\cdot]$ denotes the expectation operator.
Fig.~\ref{RWP}-(b) illustrates the RWP mobility in an attocell of radius $r_{\rm c}$. The mean of transition lengths for this attocell is $\mathbb{E}[L]=0.9054r_{\rm c}$.

\subsection{Device's Orientation}
\label{Sec:DeviceOrientation}
\begin{figure}[t!]
	\centering
	\begin{subfigure}[b]{0.45\columnwidth}
			\centering
			\includegraphics[width=1\columnwidth,draft=false]{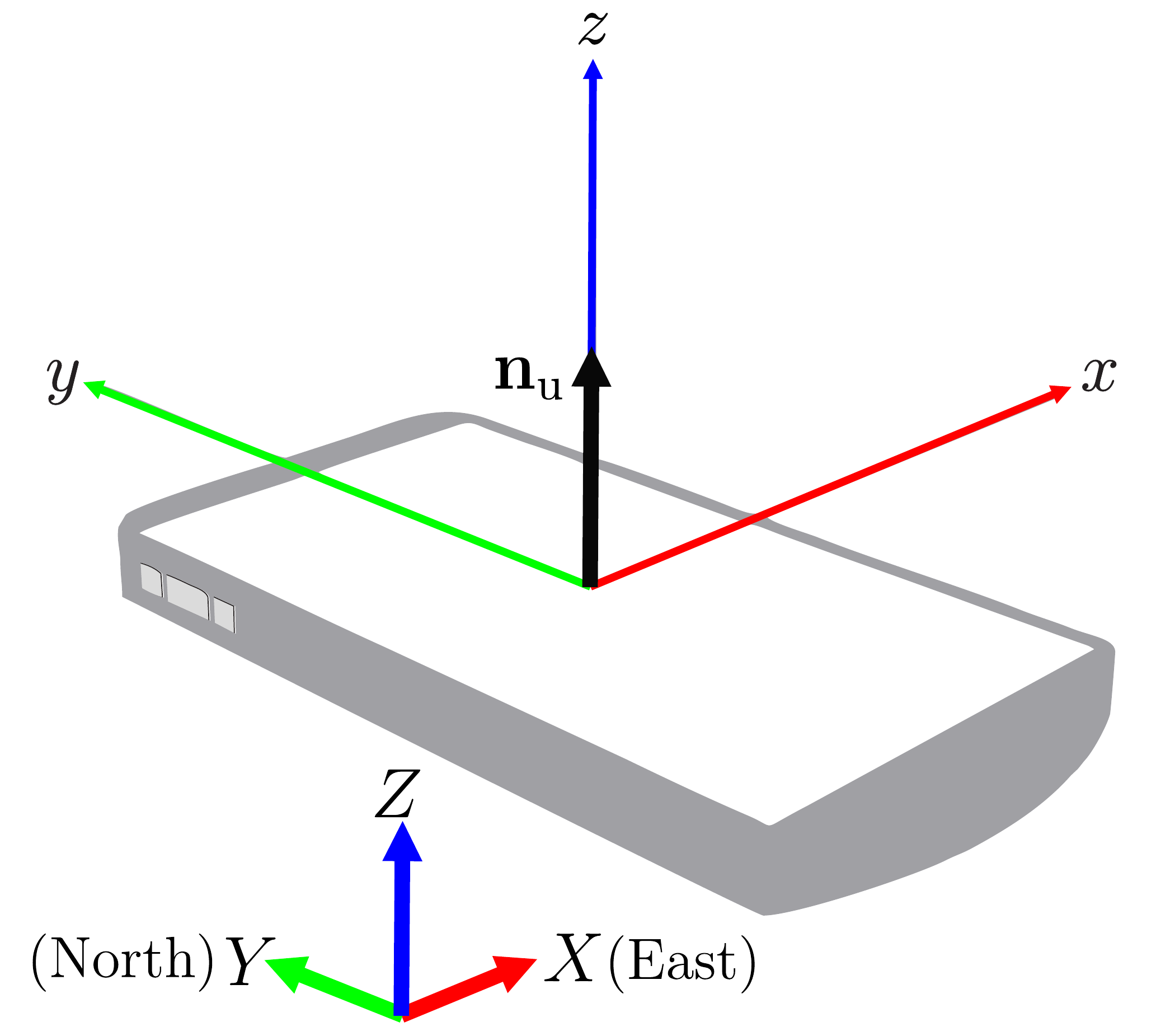}
						\caption{Normal position}
				\label{oriNorm}
		\end{subfigure}%
		~
		\begin{subfigure}[b]{0.45\columnwidth}
			\centering
			\includegraphics[width=1\columnwidth,draft=false]{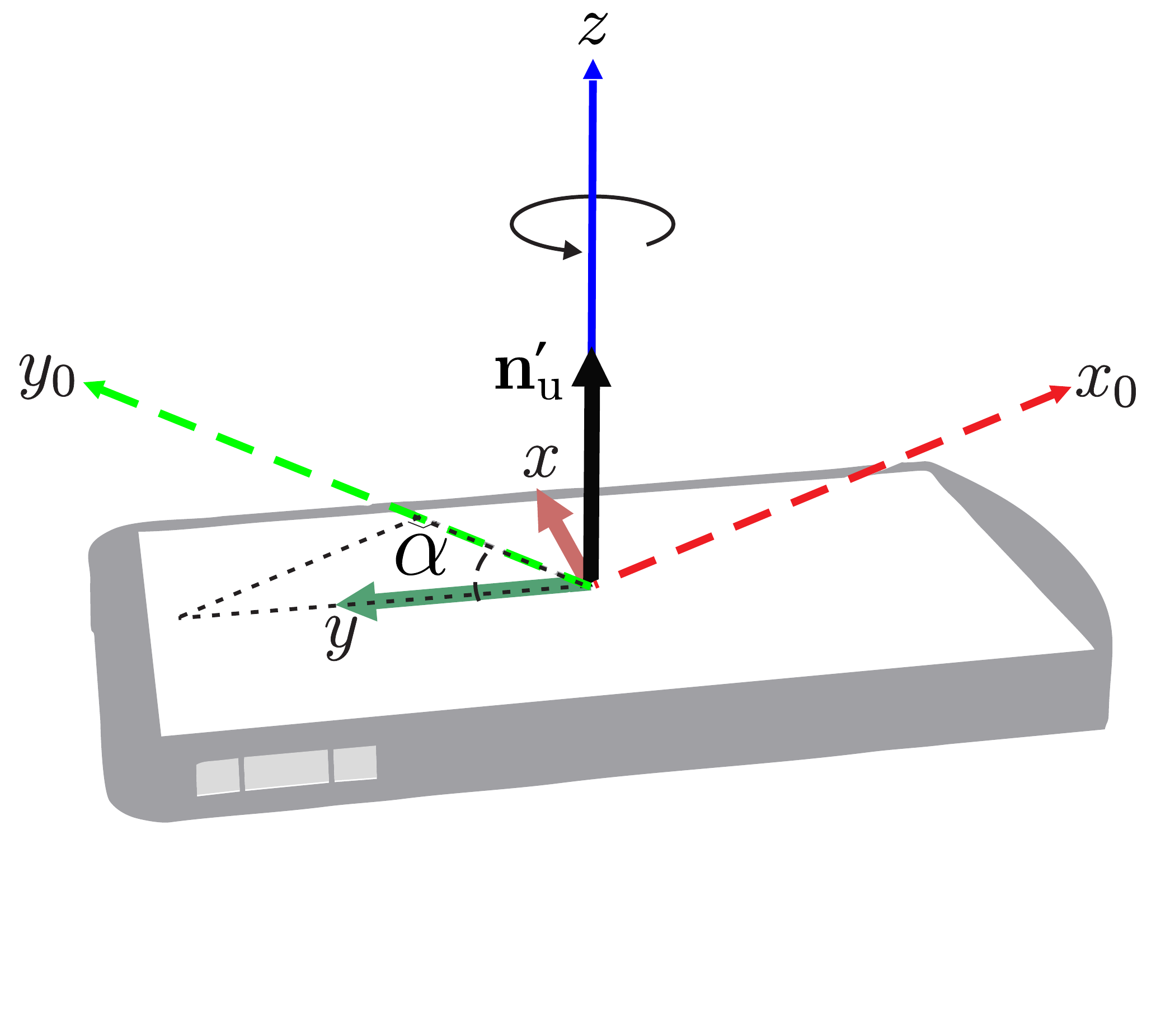}
						\caption{Yaw rotation with angle $\alpha$}
				\label{orieYaw}
		\end{subfigure}\\
	\begin{subfigure}[b]{0.45\columnwidth}
			\centering
			\includegraphics[width=1\columnwidth,draft=false]{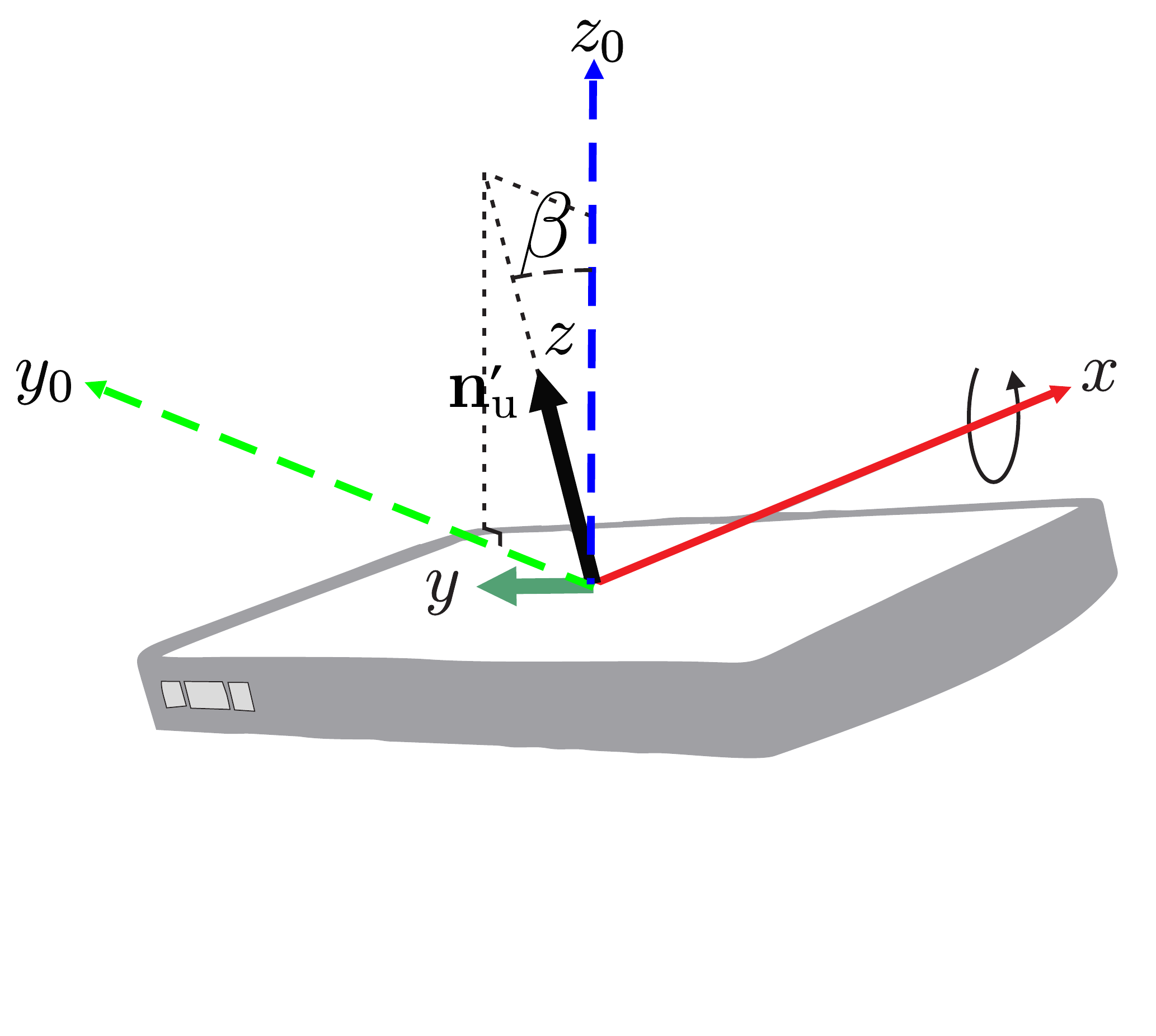}
						\caption{Pitch rotation with angle $\beta$}
				\label{oripitch}
		\end{subfigure}%
		~
		\begin{subfigure}[b]{0.45\columnwidth}
			\centering
			\includegraphics[width=1\columnwidth,draft=false]{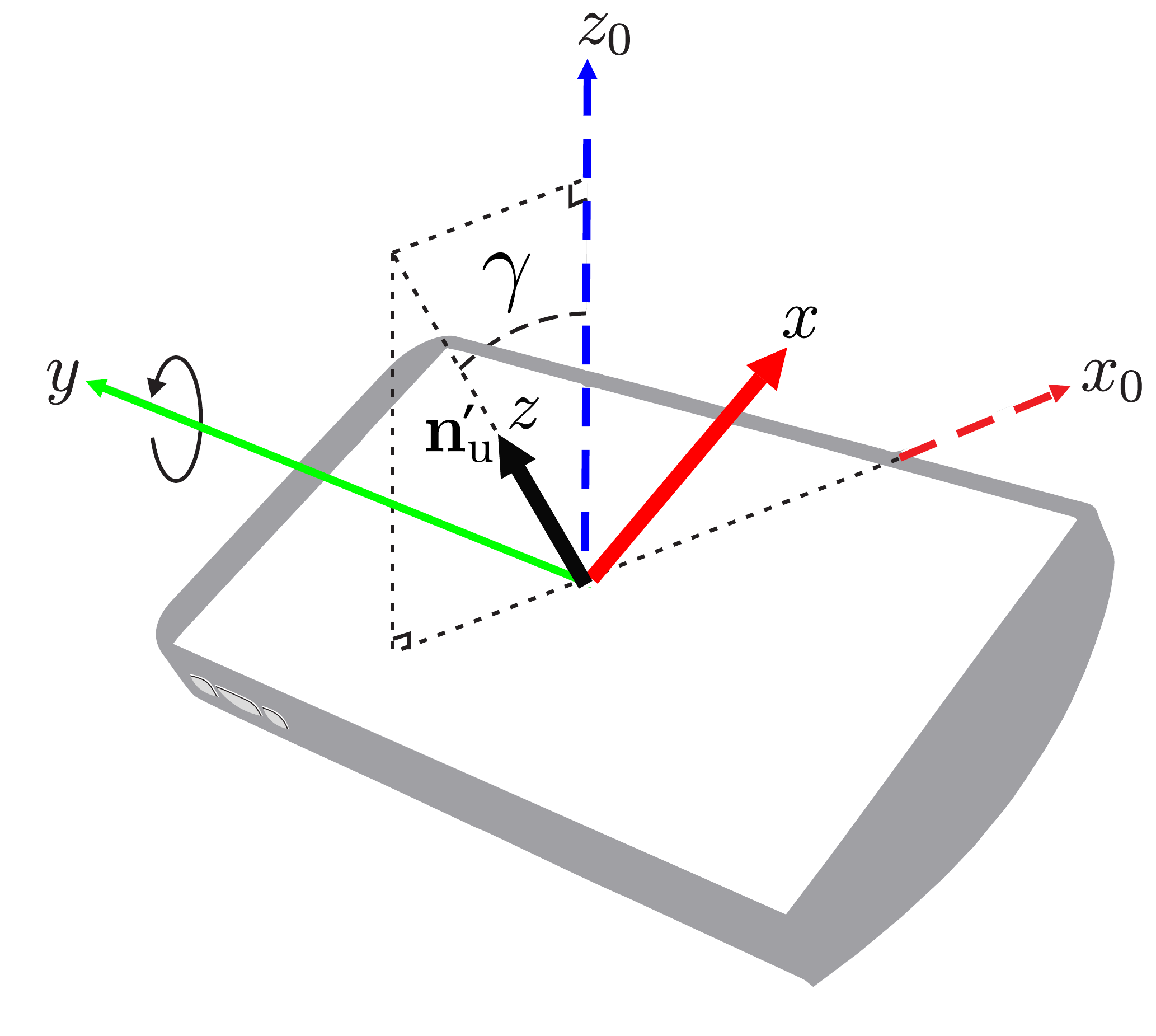}
						\caption{Roll rotation with angle $\gamma$}
				\label{orieroll}
		\end{subfigure}
	\\ 
	\caption{Orientations of a mobile device \cite{8540452}.}
	\label{FigOrientation}
	\vspace{-0.3cm}
\end{figure}

Device orientation is another factor that can affect performance of a user connected to a LiFi network in hybrid LiFi and WiFi networks. In fact, the effect of device orientation on the current WiFi systems is negligible while it is crucial in LiFi systems because of small light wavelength. 
Due to the lack of a proper model for device orientation, most LiFi-related studies have neglected the effect of random device orientation. 
Based on the Euler's rotation theorem \cite{kuipers1999quaternions}, any rotation in $\mathbb{R}^3$ space can be uniquely expressed by composing three elemental rotations, i.e., the rotations about the axes of a coordinate system. 
Thanks to the embedded-gyroscope in current smartphones, they are able to report the elemental rotation angles yaw, pitch and roll denoted as $\alpha$, $\beta$ and $\gamma$, respectively \cite{6083905}. Here, $\alpha$ represents the rotation about the $z$-axis, which takes a value in the range of $[0,360)$; $\beta$ denotes the rotation angle about the $x$-axis, that is, tipping the device toward or away from the user, which takes value between $-180^\circ$ and $−180^\circ$; and $\gamma$ is the rotation angle about the $y$-axis, that is, tilting the device right or left, which is chosen from the range $[-90,90)$. The device and Earth coordinate systems ($xyz$ and $XYZ$, respectively) are illustrated in Fig.~\ref{FigOrientation}-(a). The elemental Euler angles are depicted in Fig.~\ref{FigOrientation}-(b) to \ref{FigOrientation}-(d). The authors in \cite{7794890} used the Euler's angles to model the device orientation. Then, an AP selection algorithm with the consideration of device orientation is proposed and compared with the vertical devices and the significance of involving the orientation is confirmed. In \cite{7876858}, the impact of random device orientation is considered on load balancing in HLWNets. Their proposed load balancing algorithm can efficiently allocate resources among randomly-orientated UEs using evolutionary game theory. 

For the first time, a measurement-based model for device orientation is proposed in \cite{8540452,8377334}, where the polar angle\footnote{The polar angle is defined as the angle between the $Z$-axis and normal vector of device.} is modeled as a Laplace distribution and Gaussian distribution for sitting and walking activities, respectively. The experimental measurements of device orientation for uncontrolled activities are presented in \cite{8691024}. It is observed that the polar angle can be better fitted to a Laplace distribution in comparison with a Gaussian distribution. 
In \cite{8673906,8886097}, the impact of random device orientation on the bit-error ratio (BER) performance of a user with sitting activities in a LiFi network are evaluated. The probability density function of the signal-to-noise ratio is derived and an analytical expression for BER with the consideration of random device orientation is presented. A random process model for changes in the device orientation based on real-world measurements is introduced in \cite{8732694}. Temporal characteristics of device orientation have been considered in the random process model and the LiFi channel is modeled as a slow-varying one based on the measured coherence time. Hence, integration of random device orientation to the conventional RWP mobility model to provide a more realistic and accurate framework for analyzing the performance of mobile users is initially presented in \cite{8540452}. The performance of mobile users in LiFi networks using the orientation-based RWP (ORWP) model has been evaluated in  \cite{8790655,ChengICCW19,ImanICCW19}. All these studies use a correlated Gaussian model to generate the polar angle during the movement of a user. The statistics of the generated samples are fitted to the experimental measurements reported in \cite{8540452}. The ORWP mobility model has been used for the first time in the hybrid LiFi/WiFi network given in \cite{CSMA_HWL} to evaluate the system performance more accurately and support dynamic load balancing for mobile users.

\subsection{Light-path Blockage}
\begin{figure}[t!]
\centering
\includegraphics[width=1\columnwidth,draft=false]{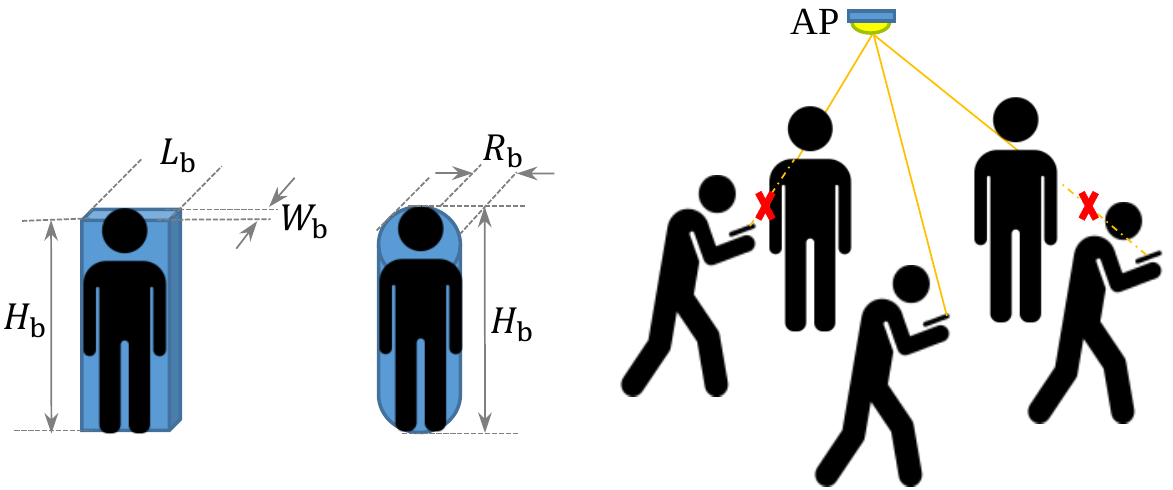}
\caption{Blockage modeling in LiFi networks.}
\label{fig:blockage}
\end{figure}

The other significant factor that should be considered in HLWNets is random blockage. It is noted that in HLWNets, LiFi systems are more vulnerable to link blockage, while WiFi systems are robust to it and the effect of link blockage is negligible. 
Due to the nature of the LiFi channel, the link between a pair of transmitters and receivers can be blocked by an opaque object such as a human body or other similar objects. However, the mobile users in the indoor environment are the major cause of link blockage. The blockers can be modeled as either rectangular prisms \cite{8471819} or cylinder objects \cite{6354257}. Fig.~\ref{fig:blockage} shows the blockage modeling for a human body either by a rectangular of size $H_{\rm b}\times L_{\rm b}\times W_{\rm b}$ or a cylinder with the height of $H_{\rm b}$ and radius of $R_{\rm b}$. 
Note that when a communication link is blocked, no extra transmission power can compensate for the error rate of the LiFi system.  
Solutions to alleviate the effect of the blockage are proposed in \cite{8790655,ImanICCW19,ChengICCW19}. A multi-directional receiver benefiting from several PDs at different sides of a smartphone is introduced in \cite{8790655,ImanICCW19} and has been evaluated in the presence of blockers and device random orientation. It is shown that the multi-directional receiver configuration outperforms the conventional structure for which all PDs are located on one side of the smartphones, for example on the screen side. In \cite{ChengICCW19}, an omnidirectional receiver is proposed, in which PDs are used on all sides of the phone to make it robust against blockage. Its superb performance is shown compared to a single-PD and two-PD configurations. 

The performance of hybrid WiFi/LiFi networks in the presence of blockers has been investigated in \cite{7056535,7511139,7876858,8464895}. 
Specifically, an analysis of LoS blockage is presented in \cite{7056535}. It is noted that when the LiFi link is blocked, the user will connect to a WiFi AP since it is capable of providing seamless coverage. In this case, the average downlink data rate will be $\bar{R}=P_{\rm b}R_{\rm wifi}+(1-P_{\rm b})R_{\rm LiFi}$ for a given blockage probability of $P_{\rm b}$; also $R_{\rm LiFi}$ and $R_{\rm wifi}$ are the data rate of LiFi and WiFi systems, respectively.
Area data rate has been improved via a hybrid WiFi and LiFi network in the presence of blockers with various densities of blockers in indoor environments \cite{7511139,7876858}. The impact of link blockage in conjunction with inter-cell interference on the user throughput in a hybrid LiFi/WiFi is evaluated. The authors proposed a load balancing algorithm based on evolutionary game theory to allocate LiFi and WiFi resources among users efficiently where the shadowing effect is considered in their proposed algorithm. It is shown that blockage is not always destructive and for low blockage density, the user data rate is high since the blockers can reduce the interference from neighboring LiFi APs. 
A simple blockage resilient hybrid LiFi and WiFi network is introduced in \cite{8464895}. The effect of LoS blockage on the LiFi performance in the hybrid LiFi/WiFi network is evaluated. It is shown that the effect of LoS link blockage can be reduced with wider LED half-intensity angles. This is due to enlarging the LiFi attocell coverage, which leads to a lower LoS blockage probability. 

\subsection{Challenges and Research Directions}
The instantaneous user behavior information has not been considered in most of HLWNets studies. This information can be fed back to the central controller by means of limited-feedback mechanisms \cite{8302445,8292542,LF2015}. One future study direction would be the utilization of this information for i) efficient allocation of the available resources to users, ii) handover management, and iii) load balancing between different technologies. No studies have investigated the impact of the availability of this information at the controller for the mentioned purposes.

Angular diversity receivers \cite{7063472,7109107,7498569,8387504,8761200}, multidirectional receivers \cite{8790655,ImanICCW19} and omnidirectional receivers \cite{8403727,ChengICCW19} have been recently developed to enhance the system performance which can be utilized in HLWNets. It is shown that these configurations are robust against the effect of blockage and device random orientation \cite{ChengICCW19,8790655}. These structures are able to improve the performance of LiFi systems by means of spatial diversity. Therefore, they can be considered as promising solutions to enhance the performance of HLWNets by offloading some amount of traffic from WiFi access point.


\section{Interference Management} \label{sec:interference_management}

\begin{figure}[t]
\centering
\includegraphics[width=3.2in]{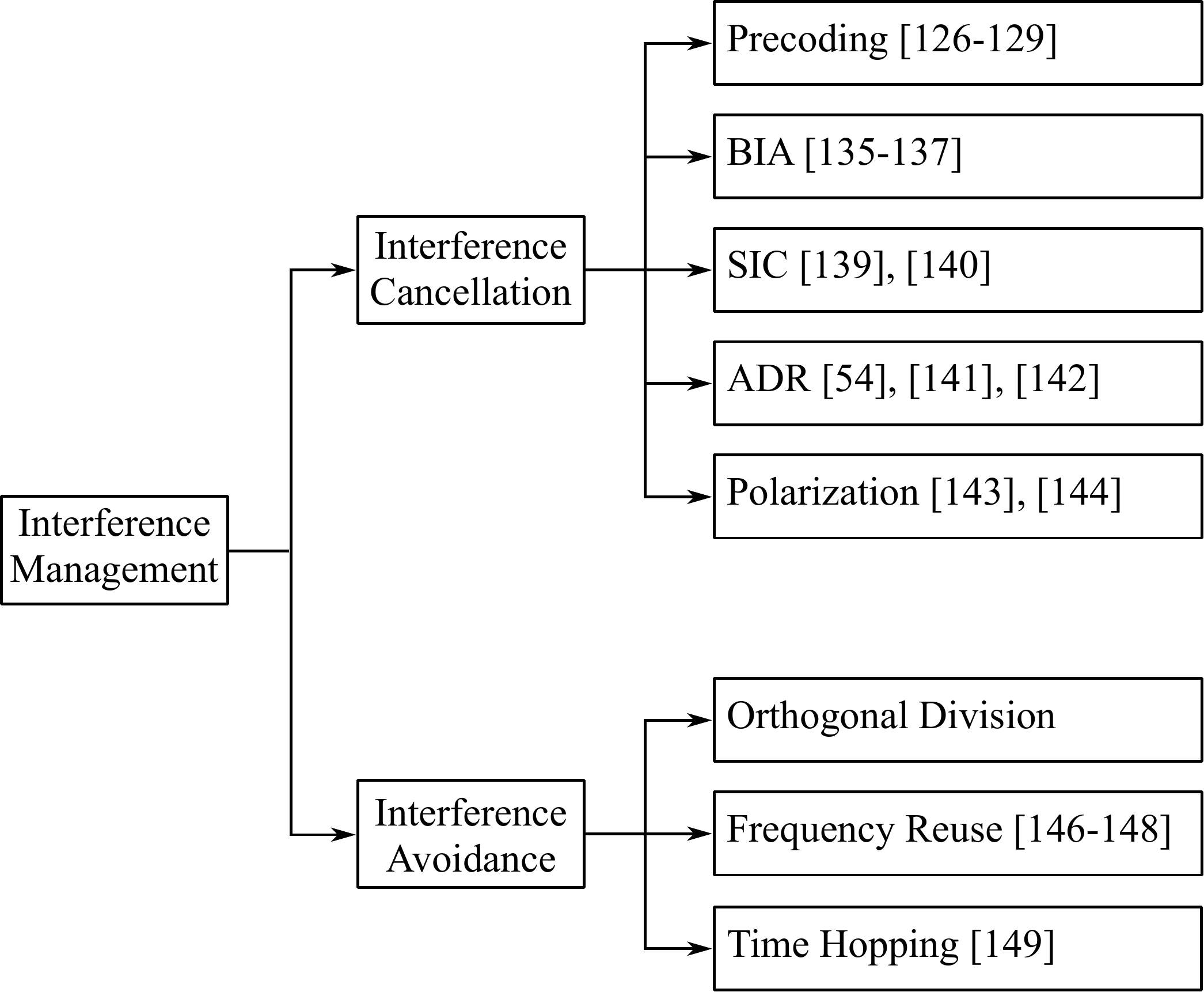}
\caption{Interference management techniques for LiFi.}
\label{fig:interference_management}
\end{figure}

Increasing the density of APs is one important aspect of network densification \cite{6736747}, which is recognized as the key mechanism for wireless evolution over the next decade, while interference management is of paramount importance. In the 3rd generation partnership project (3GPP) heterogeneous networks (HetNets) \cite{5876496}, interference management is inevitable across different network domains since they employ the same carrier frequencies. Operating at different spectra, LiFi and WiFi do not interfere with each other. In addition, the CSMA/CA adopted by WiFi can suppress co-channel interference (CCI) to a negligible level. Hence, we focus on discussing interference management techniques in LiFi. These techniques, which are summarized in \mbox{Fig. \ref{fig:interference_management}}, can be classified into two categories: interference cancellation and interference avoidance.

\subsection{Interference Cancellation}

Andrews \cite{1421925} defines interference cancellation as the class of techniques that decode desired information and utilize this information along with channel estimates to eliminate or reduce received interference from the received signal. This type of technique works at the receiver end, i.e. after the interference-affected signal is received.

\subsubsection{Precoding}
Precoding techniques (including zero forcing (ZF) \cite{1207369}, block diagonalization (BD) \cite{1677932} and dirty paper coding \cite{1512417}) are widely used to eliminate interfering signals in downlinks. The basic principle is to artificially create orthogonal channels through singular-value decomposition. Due to the non-negativity of optical signals, traditional precoding techniques need to be modified to suit LiFi, e.g. adding a DC biasing vector \cite{7106482}. Upon the type of interference dealt with, precoding techniques can be divided into two categories: multi-user detection (MUD) and coordinated multipoint (CoMP) transmission. The MUD aims to cancel interference among co-channel users within the same AP. The authors in \cite{6568880} analyzed the performance of BD precoding for LiFi with perfect CSI. Considering imperfect CSI, a precoding method based on ZF was proposed in \cite{8302455}. In \cite{7314864}, the difference in the temporal delays of optical propagation paths was considered in MUD. An optical adaptive precoding scheme was proposed in \cite{8326700}, which only nulls destructive interference where the peaks of one signal line up with the valleys of the other. The CoMP is designed to eliminates ICI, with coordination among APs being required to exchange CSI knowledge. Related research was carried out for LiFi networks in \cite{7390880,7417415,8269151}. Precoding-based interference cancellation methods require channel state information at the transmitter (CSIT) of all co-channel users. The uplink of LiFi normally employs infrared (IR) when lighting is not required \cite{7831279}, composing a frequency division duplex (FDD) system. Thus, implementing precoding techniques in LiFi comes at a cost of hefty feedback. This issue has not yet been well addressed in the current literature. Also, inaccurate CSI will impair the performance of precoding. This becomes more pronounced in LiFi, since rapid changes in the device orientation will cause fast-varying channels. The practicality of adopting precoding techniques in LiFi is still an open issue. 


\subsubsection{Blind Interference Alignment (BIA)}
When exact CSI is not available at the transmitter, the BIA can be achieved by pairing a time-selective user with a frequency-selective user \cite{6151801}. The basic concept is to obtain the maximal degree of freedom for co-channel users by masking transmitted signals on the basis of channel coherence. Unlike precoding, the BIA can only reduce interference to some extent. The condition of channel coherence for the BIA might not always be met, and thus channel manipulation is required. In RF systems, this is realized by reconfigurable antennas, which enable the receiver to switch among different antennas \cite{5734870}. A few studies have been conducted to utilize the BIA in LiFi networks. In \cite{7814294}, the BIA was applied to LiFi by equipping each user with one PD and multiple optical filters. In \cite{8417700}, the performance of the BIA was studied for HLWNets, where some users can be shifted to WiFi. However, the BIA can only outperform TDMA in the range of high optical transmit power, e.g. above 50 dBm in \cite{7814294}, which is impractical for LEDs with illumination purposes.

\subsubsection{Successive Interference Cancellation (SIC)} The SIC can detect signals that arrive simultaneously by distinguishing different power levels of the received signals. 
Illumination requirements must be complied with when power control is implemented in LiFi \cite{8527494}. Since only the AC component of the optical signal is converted to the effective electric signal, the percentage of the AC component can be adjusted, while maintaining the same average optical transmission power. In \cite{8113471}, power control was investigated when the user is served by multiple APs, with each AP consisting of multiple narrow-FoV LEDs. The SIC detects one user per stage, and thus has complexity and latency proportional to the number of co-channel users. Alternatively, parallel interference cancellation (PIC) \cite{793316} which detects all users simultaneously can reduce latency with increased complexity. In LiFi, the SIC is preferable to the PIC, since each LiFi AP covers a relatively small area and is likely to serve only a few users. Using the SIC to realize multiple access forms the concept of NOMA, which has been introduced in Section \ref{sec:NOMA}. SIC-based interference cancellation methods rely on an appropriate pair of the co-channel users, which might not always be satisfied in ultra-dense networks since there are only a few users in each cell. 


\subsubsection{Specially Tailored Methods}

There are two interference cancellation methods specially tailored for LiFi: angle diversity receiver (ADR) and polarization techniques. The ADR uses multiple narrow-FoV PDs instead of a single wide-FoV PD, in order to reduce interference at each PD. Chen \emph{et al.} \cite{8387504} analyzed the performance of different signal combining schemes for ADR, including select best combining, equal gain combining, maximum ratio combining, and optimum combining. In \cite{8471902}, the ZF precoding was combined with the conventional ADR to mitigate ISI as well as ICI. In \cite{8369104}, an optimized ADR with different numbers of PDs was proposed for different LED layouts. However, narrow-FoV PDs are susceptible to changes in the device orientation. The performance of ADR techniques is yet to be validated in a realistic mobile environment. The polarization property of light can also be used to realize differential detection for interference cancellation. A polarization-based approach was developed in \cite{7585097}, where two PDs with different polarization filters are used to cancel interference. A similar method was proposed in \cite{8305614} to resist un-polarized optical interference. This type of method requires a perfect alignment of polarization directions between the transmitter and receiver, which is feasible in laboratorial experiments but much difficult to implement in practice.

\subsection{Interference Avoidance}

The interference avoidance refers to the techniques that work at the transmitter end to avoid yielding interference. Among these techniques are orthogonal division schemes including TDMA, OFDMA and SDMA, which have been introduced in Section \ref{sec:MA}. Some studies, e.g. \cite{6157578}, list power control as an interference avoidance method. However, this type of method is unable to work without the SIC, and thus is deemed as SIC-based interference cancellation.

\subsubsection{Frequency Reuse}
Frequency reuse (FR) is widely used to avoid ICI among neighboring cells, where frequencies are reused in a regular pattern. A few studies have been carried out to apply FR to LiFi networks \cite{6155571,7163284,8064179}. In \cite{6155571}, experimental work was conducted to demonstrate the application of FR in LiFi. Fractional frequency reuse (FFR), which divides the cell area into multiple regions, was analyzed for LiFi in \cite{7163284}, including strict FFR and soft frequency reuse (SFR). The former scheme partitions the cell area into three equal sectors, while the latter one provides a two-tier cellular structure. Compared with strict FFR, SFR is more flexible and thus able to achieve a higher reuse ratio with the same capability of suppressing interference. In order to adapt to different densities of APs, a dynamic SFR scheme was proposed in \cite{8064179} with an adjustable spectrum allocation. This scheme in fact creates a more flexible cellular structure than SFR.

\subsubsection{Frequency/Time Hopping}
Frequency/time hopping methods rapidly switch a carrier among many frequency channels or time slots, using a pseudo-random sequence known to both the transmitter and receiver. A time hopping method for LiFi was reported in \cite{5277960}, where the period and duty cycle of the optical carrier are varied in a pseudo-random manner. This type of method reduces the probability of two users occupying the same time-frequency block, but requires a strict synchronization between the AP and user.

\subsection{Challenges and Research Directions}

User mobility is a critical issue to LiFi, due to the relatively small coverage area of a single LiFi AP. However, the impact of user mobility on interference management has not been well addressed in the current literature. In contrast to interference cancellation, interference avoidance is more tolerant to user mobility. Among these methods, FR schemes are a promising interference management technique for LiFi, since they can effectively utilize the huge optical spectra. However, how to implement FR schemes in LiFi is still an open issue. There are two ways, depending on using a single wavelength or multiple wavelengths. When a single wavelength is adopted, the baseband bandwidth of same color LEDs are divided for FR, and each AP uses only a portion of the bandwidth. However, the bandwidth of a single LED is limited due to its size and material. The modulation bandwidth of commercial LEDs is usually 10-20 MHz \cite{cossu2012}. Therefore, using a single wavelength will severely restrict the modulation bandwidth of each AP. 

With multiple wavelengths being employed for FR, the full bandwidth of each color LED can be exploited. The problem is how to distinguish the signals carried by different wavelengths. As we know, PDs accept a range of wavelengths, depending on the materials used. Most widely used silicon PDs are sensitive from 400 to 1100 nm, covering almost the entire visible spectra. Within this accepted range, PDs convert all of the received photons into an electric current, and thus are unable to separate the signals of different wavelengths. A straightforward solution is to place an optical filter before the PD, e.g. in \cite{8535210}. Adaptive optical filters are essential for receiving variable wavelengths. In \cite{6056538}, a spectrum sensor array was proposed to dynamically select the signals carried by different wavelengths. The area of such an array is proportional to the number of supported wavelengths. More research is needed to develop a compact and efficient design for adaptive optical filters, e.g. using glasses that change the attenuation based on intensity.

Also, WiFi can be used to mitigate LiFi interference in order to fully exploit the advantages of HLWNets. Specifically, the users causing severe interference in LiFi can be shifted to WiFi. For instance, the involvement of WiFi could help pairing co-channel users effectively in NOMA. A research gap exists in implementing network-interactive interference management in HLWNets.



\section{Handover} \label{sec:handover}

In mobile communications, a handover occurs when the user with a data session is transferred from one AP to another without disconnecting the session. According to the sequence of break and connection, handovers are classified into two categories: hard handovers and soft handovers. With a hard handover, the user first breaks its current connection and then builds a new connection to another AP, referred to as a target AP. During this handover process, the user and the target AP exchange signalling information (termed handover overhead) via dedicated channels and no data transmission is available. As for a soft handover, a new connection is made before the user is disconnected from the host AP. This allows data transmission in parallel with the handover process. Normally soft handovers can be used when the host and target APs operate within the same carrier frequency. On the contrary, a handover that needs a change in the carrier frequency is performed as a hard handover. Therefore, it is necessary to study how the handover process affects the network performance of HLWNets.

In general, the handover process in a hybrid network falls into two categories: horizontal handover (HHO) and vertical handover (VHO). A HHO takes place within the domain of a single wireless access technology, whereas a VHO occurs between different technologies. With a VHO, the air interface is changed, but the route to the destination remains the same. In some literature, e.g. \cite{6587998}, a third category named diagonal handover is also introduced, with the air interface and route to the destination both being changed. A significant body of research was conducted in the field for HetNets, and a related survey was carried out in \cite{7462492}, which summarizes received signal strength (RSS)-based, load balancing-related, and energy-saving handover schemes. However, handling the handover process is more challenging in HLWNets than in cellular HetNets.


\subsection{Horizontal Handover (HHO)}

HLWNets have a limited coverage range with a single AP, especially with a LiFi AP covering 2-3 meters in diameter \cite{7121823}. The ultra small cell makes HLWNets encounter considerably frequent handovers, even when the user moves at a moderate speed. Also, it is worth noting that the LiFi channel is related to the PD's receiving orientation. The alteration of the PD's receiving orientation could be very rapid and sudden, leading to frequent and unexpected handovers. For the above reasons, the handover overhead becomes a critical factor affecting the network throughput of HLWNets.

Taking handovers into account, the separation distance between APs affects network throughput in two aspects. On the one hand, a smaller separation can provide a higher area spectral efficiency. On the other hand, a larger separation can decrease the handover rate. Motivated by this, the optimal placement of LiFi APs was studied in \cite{7328268}, which concludes that the optimal coverage area of a LiFi AP is 2 to 8 $\text{m}^{2}$, depending on the user density and handover overhead. While the coverage areas of different APs normally overlap each other, the authors in \cite{6167395} also investigated the handover procedure for non-overlapping coverage. This study suggests a soft handover for non-overlapping coverage and a hard handover for overlapping coverage. However, the above two papers only involve user mobility and fail to consider the rotation of mobile devices. In \cite{7925676}, the handover rate was analyzed in a LiFi network with both the movement and rotation of user equipment being considered. It is found that the handover rate peaks when user equipment is tilted between 60\textdegree{} and 80\textdegree{}.

Although the optimal placement of APs can relieve the detriment of handovers to some extent, the degradation in throughput is still outstanding for fast-moving users. The concept of handover skipping (HS) was introduced in \cite{7792669}, which enables the user to be transferred between non-adjacent APs. In this work, a topology-aware HS scheme was proposed to let the user skip some APs if the chord length of the cell is below a pre-defined threshold. A similar method was reported in \cite{8304530}, but with the research scope being extended to multi-AP association. This type of method relies on knowledge of the user's trajectory and network topology. However, the equivalent network topology of LiFi is dynamic and user-dependant, due to the impact of the PD's receiving orientation. Also, positioning techniques are required to acquire knowledge of the user's trajectory and uplink transmission is needed to feed this information to APs. To circumvent the above stringent requirements, an RSS-based HS approach was developed in \cite{8662599}. This method exploits the rate of change in RSS to reflect whether the user is moving towards a certain AP. A weighted average of RSS and its rate of change is formed to determine the target AP for handover. Since RSS is commonly used in the current handover schemes, no additional feedback is required. More importantly, this HS approach does not rely on knowledge of the network topology. It is claimed that the HS method in \cite{8662599} can improve user throughput by up to about 70\% and 30\% over the LTE handover scheme \cite{3gpp_r12} and the trajectory-based HS method, respectively.

\subsection{Vertical Handover (VHO)}

The user usually requires a VHO from LiFi to WiFi when losing LiFi connectivity. The loss of LiFi connectivity might be caused by two reasons: i) the light-path is blocked by opaque objects, such as human bodies and furniture; and ii) the PD's receiving orientation is significantly deviated from the LoS path. In \cite{8417682}, the probability of VHO was analyzed for a user with random rotations in the HLWNet. Studies were also carried out to develop VHO schemes for LiFi-involved hybrid networks \cite{7367915,7390808,7827135}. In \cite{7367915}, a VHO scheme based on the Markov decision process was proposed. This method determines whether to perform a VHO on the basis of the queue length for WiFi and the channel condition of LiFi. Another VHO scheme was proposed for hybrid LiFi and LTE networks in \cite{7390808}, which predicts the system state in terms of interruption durations, message sizes and access delays. These parameters, recorded by the user equipment in real time, are used to make handover decisions. A similar approach was developed for hybrid LiFi and femto networks in \cite{7827135}, which considers multiple attributes including dynamic network parameters (e.g. delay, queue length and data rate) and actual traffic preferences. However, without weighing the advantages and disadvantages of VHO and HHO, the above methods are unable to provide the optimal solution.

\subsection{A Choice between HHO and VHO}

\begin{figure}[!t]
\centering
\includegraphics[width=3.2in]{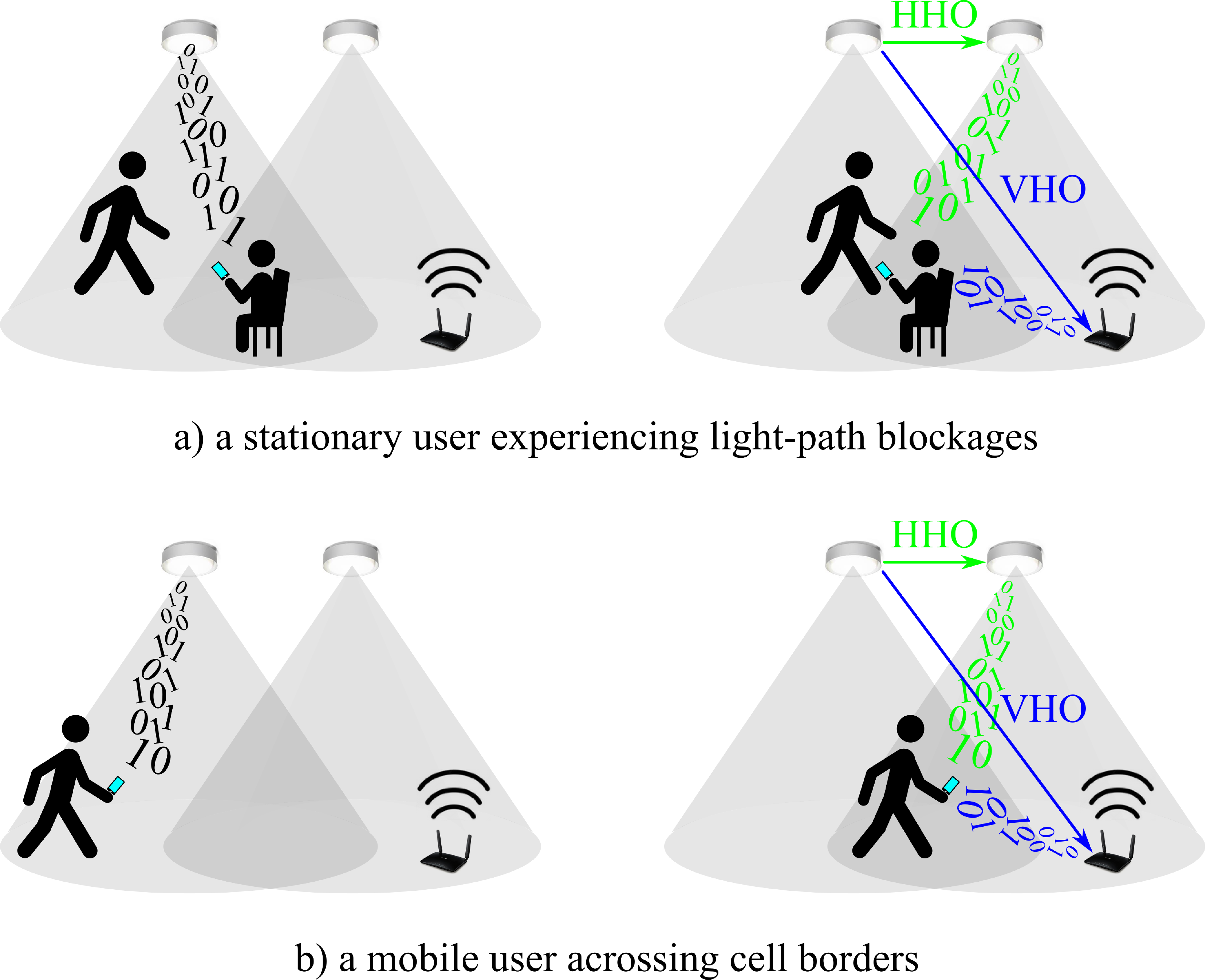}
\caption{Examples of choosing between HHO and VHO.}
\label{fig:HHO_and_VHO}
\end{figure}

Due to the change of air interfaces, a VHO usually needs a much longer processing time than a HHO \cite{3GPP_25913}. Also, the WiFi system has a lower system capacity than LiFi, and an excessive number of WiFi users would cause a substantial decrease in throughput. Thus, the choice between HHO and VHO is critical to HLWNets, which is exemplified in Fig. \ref{fig:HHO_and_VHO}. Specifically, not all of the users losing LiFi connectivity should be switched to WiFi, e.g. the users encountering a transient light-path blockage. Apart from that, the user's velocity is also an important factor in deciding whether a user should be served by LiFi or WiFi. In general, fast-moving users prefer WiFi, since they would experience frequent HHOs in LiFi. To solve this complicated problem, Wang \emph{et al.} \cite{7510823} proposed a handover scheme based on fuzzy logic. This method makes the handover decision by measuring parameters including not only CSI, but also the user's speed and data rate requirement. But this method fails to address the issue of light-path blockages. By exploiting the statistical information on light-path blockages, the authors in \cite{8227704} formulated the handover process as an optimization problem to maximize network throughput over a period of time. The issue of choosing between HHO and VHO essentially involves load balancing, which will be elaborated in Section \ref{sec:load_balancing}.

\subsection{Challenges and Research Directions}

With respect to the handover topic in HLWNets, most current literature considers the user to be served by a single AP at a time, i.e. single-AP association. Due to the dense deployment of APs and intermittent light-path blockages, frequent handovers could occur despite the techniques that aim to minimize the impact caused by handovers. Meanwhile, multiple-AP association, which allows the user to receive parallel transmissions from multiple APs, has been widely ignored. This is because the conventional transmission control protocol (TCP) does not support the packets sent from different APs to be reordered at the destination. Since 2013, the Internet Engineering Task Force (IETF) has been working on the multipath transmission control protocol (MPTCP) \cite{multipath_TCP}, which adds a subflow sequence number in the packet overhead to solve the issue of packet reordering. Adopting multiple-AP association could potentially improve the network performance of HLWNets by avoiding handovers without losing connectivity. For example, a user experiencing light-path blockages can still be served by WiFi without the need for handover. Future research needs to focus on developing handover schemes in multiple-AP association scenarios, to achieve a truly seamless handover in HLWNets.

\section{Load Balancing} \label{sec:load_balancing}

In the field of wireless networks, load balancing (LB) refers to the techniques that distribute user sessions across APs with overlapping coverage areas. The aims of LB techniques are to optimize resource utilization, to maximize throughput, to minimize response time, and to reduce network congestion. In homogeneous networks, the coverage overlap is restricted among APs to mitigate ICI. As a result, LB only applies to cell-edge users when they impose unbalanced traffic loads to different APs. In other words, LB is not needed when the users' demands for data rates are uniformly distributed in geography. The authors in \cite{4747629} conducted an overview of WiFi-related LB techniques, which are classified into two categories: user-based and AP-based. With user-based methods, each user selects APs according to its own interest, with the optimal network performance being hardly achieved. In contrast, AP-based methods implement network-wide LB but require a central unit to coordinate APs.

Unlike in homogeneous networks, LB becomes essential and challenging in hybrid networks including HLWNets. This is because the coverage areas of LiFi and WiFi completely overlap each other. Also, WiFi APs have a larger coverage area but a lower system capacity than LiFi APs \cite{7362097}. This renders WiFi susceptible to traffic overload even if the users' demands for data rates are uniformly distributed in geography. A large body of research was carried out to study LB approaches in HetNets, including: relaxed optimization, Markov decision process, game theory and cell range expansion \cite{6812287}. Though they are applicable to HLWNets, these methods face the critical issue of user mobility due to the short coverage range of a single LiFi AP. In the existing literature, the LB algorithms developed for HLWNets, which are summarized in Table \ref{table:LB_approaches}, can be classified into two categories: i) stationary-channel and ii) mobility-aware.

\begin{table*}[t]
\renewcommand{\arraystretch}{1.2}
\caption{A summary of LB approaches in HLWNets.}
\label{table:LB_approaches}
\centering
\begin{tabular}{|l|l|l|l|l|l|l|}
\hline
Algorithm type & \vtop{\hbox{\strut Mobility}\hbox{\strut awareness}} & Ref. & Required knowledge & Fairness & Complexity & Note \\
\hline
\multirow{4}{*}{\vtop{\hbox{\strut Global}\hbox{\strut optimization}}} & \multirow{2}{*}{N} & \cite{7056535} & CSIT & PF & High &  \\
\cline{3-7}
  &  & \cite{8458229} & CSIT & PF & High & Consider different data rate requirements \\
\cline{2-7}
 & \multirow{2}{*}{Y} & \cite{8227704} & CDT statistics & PF  & Medium & \\
 \cline{3-7}
  &  & \cite{7925839} & CDT and blockage statistics &   PF & Medium & Consider intermittent light-path blockages\\
\hline
\multirow{5}{*}{\vtop{\hbox{\strut Autonomous}\hbox{\strut optimization}}} & \multirow{2}{*}{N} & \cite{7876858} & CSIT & MFF & Medium & Consider different data rate requirements \\
\cline{3-7}
 &  & \cite{8357657} & CSIT & PF & Medium & Consider power allocation \\
\cline{2-7}
 & \multirow{3}{*}{Y} & \cite{7536188} & CSIT and users' trajectories & PF & Medium & Use the college
admission model \\
 \cline{3-7}
  &  & \cite{7274270} & CSIT & MFF & Medium & \\
 \cline{3-7}
  &  & \cite{7820066} & CSIT & MFF & Medium & \vtop{\hbox{\strut Compare joint and separate implementations}\hbox{\strut of AP assignment and resource allocation}} \\
\hline
\multirow{2}{*}{\vtop{\hbox{\strut Direct}\hbox{\strut decision-making}} } & N & \cite{8013858} & CSIT & PF &  Low & Use fuzzy logic \\
\cline{2-7}
 & Y & \cite{7510823} & CSIT and users' speeds & PF & Low & Use fuzzy logic\\
\hline
\end{tabular}
\end{table*}

\subsection{Stationary-channel Load Balancing}

The issue of LB in HLWNets was studied for stationary channels in \cite{7056535,8458229,7876858,8357657,8013858}. In \cite{7056535}, an LB method was proposed to achieve proportional fairness (PF) among users, in forms of both centralized and distributed resource-allocation algorithms. To improve quality of service, the LB issue is formulated as a mixed-integer non-linear programming problem in \cite{8458229}, which considers different data rate requirements among users. The two above methods both form an NP-hard problem, and solving the problem requires a computational complexity that exponentially increases with the number of APs. To reduce processing power, an iterative algorithm based on evolutionary game theory was reported in \cite{7876858}, with multiple fairness functions (MFF) being considered. In this work, light-path blockages, arbitrary receiver orientations and data rate requirements are characterized to model a practical communication scenario. The authors in \cite{8357657} also introduced an iterative algorithm but focused on power allocation. This algorithm consists of two states: i) finding the optimal power allocation of each AP to maximize its throughput; and ii) seeking another AP for the user with the minimum data rate to increase the overall throughput and to enhance the system fairness. The above iterative algorithms, which require quantities of iterations to reach a steady state, can be deemed as autonomous optimization.

As mentioned, global optimization and autonomous optimization have their respective limitations, resulting in a considerable amount of processing time. In HLWNets, CSI could rapidly vary for mobile users with an even modest speed. This restricts the processing time, and thus challenges the practicability of the above methods. Alternatively, direct decision-making methods are applicable, which provide a significantly reduced amount of processing time. Such an LB method was reported in \cite{8013858}, which firstly determines the users that should be served by WiFi and then assigns the remaining users as if in a stand-alone LiFi network. Relying on statistical knowledge of data rate requirements and CSI, this fuzzy logic-based method is able to achieve near-optimal performance in terms of throughput and user fairness, while reducing the processing time by over 10 orders of magnitude.

\subsection{Mobility-aware Load Balancing} \label{sec:malb}

The methods noted in the previous section share a common property: relying on knowledge of CSI. While CSI varies due to user movements or environmental changes, these methods have to recalculate their solutions accordingly. When the new solutions make a change to the situation of user association, which might change very rapidly in HLWNets, the impact caused by handovers must be considered. For instance, users at the cell edge of a LiFi AP would prefer WiFi and users at the cell center of a LiFi AP would prefer LiFi. With stationary-channel load balancing methods, users would be transferred between LiFi and WiFi repeatedly when moving across the LiFi APs.

In order to tackle the above issue, user mobility needs to be considered in conjunction with LB. This is referred to as mobility-aware load balancing. In \cite{7536188}, such an approach was proposed on the basis of the college admission model. Specifically, the achievable data rate and the user's moving direction are used to measure the user's preference, and the sum data rate of the served users is exploited to compute the AP's preference. These two preferences are then iteratively calculated to reach a steady solution. A dynamic LB scheme was proposed in \cite{7274270}, which also performs an iterative algorithm. In each iteration, AP assignment and RA are sequentially implemented to improve the effective data rates which excludes handover overheads. In \cite{7820066}, a similar method was developed but with a joint implementation of AP assignment and RA. Results show that a joint implementation can achieve 50\% more throughput than a separate implementation, at the cost of a higher computational complexity by 3 orders of magnitude. In \cite{7925839}, a globally-optimized LB method is realized by using cell dwell time to measure the handover cost. All the above methods have one common point: they trade off the instantaneous data rate with the handover rate to maximize the average data rate.

\subsection{Challenges and Research Directions}

User mobility affects LB in two aspects: i) the handover cost and ii) the response time, which consists of two components: i) the processing time of the LB algorithm and ii) the round-trip time of exchanging information between the control unit and the AP. In order to support user mobility, LB methods must have a sufficiently short response time. While the current mobility-aware LB methods take the handover cost into account, they have only been examined offline via simulations. The feasibility of these methods in a real-time system is yet to be tested. This test will possibly take place on an SDN platform. 

Arbitrary receiver orientations and light-path blockages are widely ignored in the existing LB methods designed for HLWNets. Arbitrary receiver orientations would alter the coverage areas of LiFi APs \cite{MDSPIMRC19}, while light-path blockages would raise a request for dynamic LB. Though light-path blockages are considered in \cite{7876858} and \cite{7925839}, these papers use a simple model which is not established upon a realistic scenario. Also, how arbitrary receiver orientations will affect the performance of HLWNets is still an open question.

With respect to data rate requirements, the current studies consider either none or fixed ones. In practice, the data rate requirement of a user varies with time and fits a log-normal distribution \cite{8737483}. A time-varying data rate requirement would cause unbalanced traffic loads in different time slots. This factor further challenges LB methods. Burst transmission or discontinuous transmission techniques could be associated with the conventional LB methods to address this under-researched issue.

\section{Applications} \label{sec:applications}

Apart from serving communication purposes, HLWNets can also be utilized to provide or enhance some application services. In this section, we overview the use of HLWNets in two application areas: indoor positioning system (IPS) and physical layer security (PLS).

\subsection{Indoor Positioning} \label{sec:positioning}

Positioning is an essential tool for providing location-based services such as navigation and creating maps and tracking objects, etc. As the mainstream positioning technology at present, the global positioning system (GPS) is a satellite-based radio-navigation system that provides geolocation to a GPS receiver anywhere on Earth \cite{GPS}. With the lastest stage of accuracy enhancement using the L5 band centered at 1176.45 MHz, GPS can improve its accuracy from \mbox{5 m} to 30 cm \cite{GPS_accuracy}. However, GPS becomes less accurate in indoor scenarios, because the transmitted signals are degraded and interrupted by obstructions, especially ceilings and walls. Alternatively, IPS is developed on the basis of short-range wireless communication technologies, e.g. WiFi \cite{4266950}, LiFi \cite{positioning_VLC_survey_1}, Bluetooth \cite{7103024}, radio frequency identification \cite{5953513} and ZigBee \cite{6156509}. Multiple surveys have been carried out to summarise LiFi-based positioning techniques \cite{positioning_VLC_survey_2,8015106,8292854}. An overview of WiFi-based positioning methods was reported in \cite{7060497}. In this section, a brief introduction on the development and classification of IPS techniques (especially those based on LiFi) is given. Our focus is to shed light on positioning in HLWNets by comparing the IPS techniques.

\subsubsection{Classifications of IPS Techniques}

\begin{figure}[ht]
\centering
\includegraphics[width=3.2in]{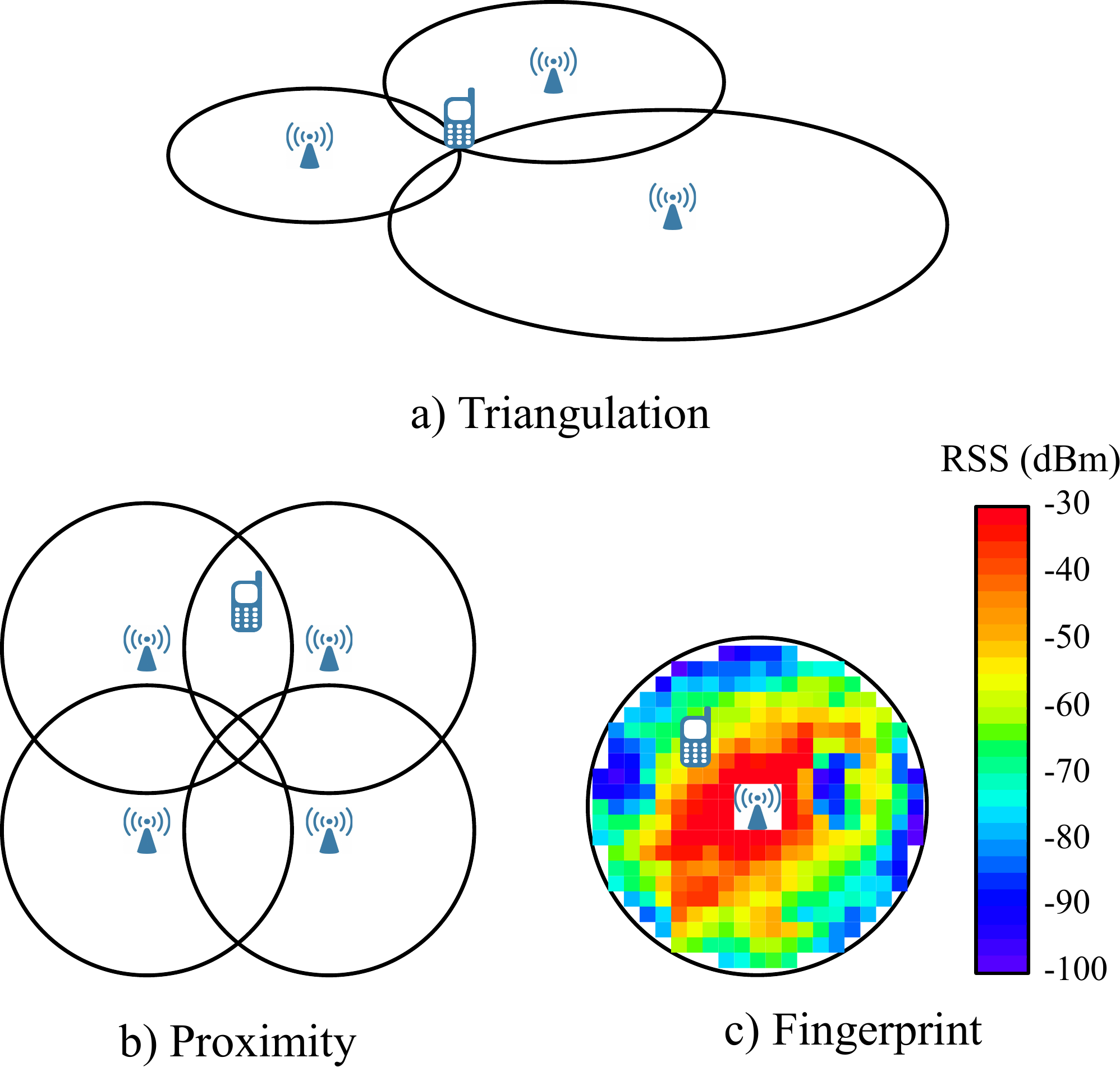}
\caption{Different positioning methods: a) Triangulation b) Proximity and c) Fingerprint.}
\label{fig:positioning_methods}
\end{figure}

In general, IPS approaches can be classified from two perspectives: the mathematical method and required information. In the current literature, there are three types of mathematical methods used for IPS: triangulation, proximity and fingerprint \cite{8015106}, as shown in Fig. \ref{fig:positioning_methods}. Among these methods, triangulation is the most popular algorithm, which uses the geometric properties of triangles by measuring the distance or angle from the device to fixed known points (named beacons)  \cite{7368080,6600748,6131130,6868970}. Triangulation is usually the most accurate, but also highly complex in terms of facility and computation. With a single receiver, at least three APs are required for a 2-D location measurement and four APs for 3-D. Proximity is the simplest positioning method, where the location of a mobile device is associated with the coverage range of nearby APs \cite{6577276}. If the device is identified by multiple APs, the location can be estimated within the overlapping coverage areas of those APs. As a result, this type of method has relatively poor accuracy if the overlapping area is large. Compared with proximity, fingerprint can provide higher accuracy by using location-dependant information (e.g. RSS), but requires off-line radio-maps \cite{fingerpring_vlc}. The optimal location is obtained by minimizing the Euclidean distance between the information in the radio-map and the online measurement \cite{shi2018accurate}. Accordingly, the accuracy of fingerprint is dependent on the radio-map.

According to different required information, IPS techniques fall into four categories: RSS, time of arrival (TOA), time difference of arrival (TDOA) and angle of arrival (AOA) \cite{8292854}. RSS-based methods, e.g. \cite{7368080}, exploit channel attenuation to estimate the distance between the device and beacon. When compared to other characteristics of the transmitted signal, the RSS can be readily acquired. The accuracy of such a method depends on a reliable path-loss model, and might suffer from uncontrollable errors caused by multipath propagation, i.e., the small-scale fading. TOA-based methods, e.g. \cite{6600748}, also compute the distance but use the travel time of the signal multiplied by the speed of light. This type of method needs rigid time synchronization between the device and beacon. To avoid this requirement, TDOA-based methods employ multiple transmitters or receivers to obtain the time difference between transmitted signals \cite{6131130}. However, time synchronization is still required between beacons. AOA-based methods, e.g. \cite{6868970}, measure the angle between the transmitted signal and the normal angle of the beacon. In RF, AOA is usually obtained by detecting the phase difference between antennas \cite{5571900}. However, AOA cannot be measured directly in LiFi due to the lack of phase information in IM/DD. Instead, AOA can be acquired through two approaches. One approach is called image transformation \cite{image_transformation}, which calculates AOA by using the trigonometric relationship between the light beacons' coordinates and their imaging locations on a photo. The other one is modeling \cite{6503631}, which exploits the angular pattern of RSS at the PD.

\subsubsection{IPS in Hybrid Networks}

Accuracy is the key metric for positioning systems. The intrinsic shorter coverage range of LiFi leads to a smaller positioning error (0.1-0.35 m) when compared to WiFi (1-7 m) \cite{positioning_VLC_survey_1}. Also, LiFi can provide more dense beacons than WiFi. Due to the existing and ubiquitous lighting infrastructures, the installation cost and energy consumption of LiFi-based IPS are low. Furthermore, LiFi-based IPS can readily detect the device's orientation via angle diversity receivers \cite{7335563}, whereas it is difficult for IPS based on RF. Moreover, there are different challenges in propagation for IPS based on LiFi and WiFi. WiFi signals might experience severe multipath effects, especially in some challenging environments such as undergroud mines and tunnels. While LiFi signals could face light-path blockages caused by opaque objects, resulting in a complete loss of connectivity.

Therefore, a hybrid IPS using both LiFi and WiFi (or other RF technologies) is envisaged to improve the performance of IPS, and a few attempts have already been made \cite{6414983,6173463,8211505}. In \cite{6414983}, the proximity positioning concept was applied in a hybrid environment of LiFi and Zigbee. This method however can only detect a 2-D position, with a relatively low accuracy ($<$130 cm). In \cite{6173463}, a two-stage positioning system was proposed. This system first determines a possible position area via a LiFi-based proximity method, and then locates the specific position in the possible area by using the RSS of the RF signals to minimize error. Such a system is able to keep the position estimation error within 20 cm. Another two-stage positioning system was developed in \cite{8211505}, but using the opposite order of RF and LiFi. In the first stage, RF is used to detect which room the device is currently located in. In the second stage, LiFi is employed to locate the device within the room. This system was reported to attain an estimation error of 5.8 cm. Nonetheless, the above two studies are still focused on detecting 2-D positions.

\subsection{Physical Layer Security (PLS)}

In wireless communications, the signals are transmitted in open air and can be received by the intended user as well as the eavesdropper (referred to as Bob and Eve). The PLS technologies can be classified into two categories: key generation scheme and transmission scheme. Typically, the method of secure key generation is based on the inherent randomness of the wireless channel by exploiting the channel characteristics \cite{ren2011secret}. For example, the secure key can be extracted based on variable RSS or channel phase. With well-designed transmission schemes, PLS aims to enlarge the SINR difference between the links of Bob and Eve. Usually, Eve's SINR performance can be weakened in two ways: i) reducing RSS and ii) increasing noise/interference \cite{8509094}. The first way is focused on optimizing the transmission scheme for Bob through techniques such as beamforming, resource allocation and interference alignment, etc. This aspect is in line with the interference management. As a result, the transmission power can be reduced for Bob as well as for Eve. The second way is to inject artificial noise, but this type might deteriorate Bob' SINR which becomes sensitive to channel estimation errors. The noise can be generated in the null-subspace of Bob's channel so that only Eve is impaired by the noise. Also, when Eve's channel is worse than that of Bob's on average, secure channel coding such as low-density parity-check \cite {4276938} can effectively increase secrecy. 

Compared with WiFi, LiFi has intrinsic security advantages. Firstly, light does not penetrate opaque objects, and hence LiFi can be securely used in a compartment. The typical scenarios pluralise this list, i.e., libraries, classrooms and office room, etc. Secondly, LiFi covers a shorter range than WiFi with a single AP, meaning Eve needs to be closer to the transmitter. Thirdly, the transmission power of LiFi concentrated on LoS is usually above 85\% \cite{7362097}. As a result, there would be very limited information leakage to Eve without a LoS path. A quantity of research \cite{arfaoui2019physical,7106457,7996639,8119998,8272327,mostafa2015physical} has been conducted to analyze the secrecy performance of LiFi, which is commonly measured by secrecy rate/capacity, i.e., the difference between the channel capacities of Bob and Eve. Specifically, Chen \emph{et al.} \cite{7996639} compared different types of LiFi deployments including: hexagonal, square, PPP and hard-core point process. It was proved that the hexagonal deployment provides the highest secrecy rate, while square performs marginally worse. Ayman \emph{et al.} \cite{mostafa2015physical} considered the secrecy capacity of an amplitude-constrained Gaussian wiretap channel, and the beamforming was utilized for the multiple-input single-output (MISO) wiretap channel. In addition, quantum key distribution (QKD) is a specific application for PLS technology in optical communication \cite{humble2013quantum}, since a photon can be easily encoded as zero or one state, such as the horizontal and vertical polarization state, long path and short path state, etc. The advantage of QKD is the unconditional security against unbounded Eve since its security is managed by quantum mechanics such as the quantum no-cloning principle.

With respect to HLWNets, a few studies have also been carried out on the topic of PLS. The secrecy performance of the RF uplink was analyzed in \cite{7935472}, with light energy harvesting being considered in the LiFi downlink. The work in \cite{7986296} aims to minimize the power consumption of the HLWNet, while satisfying the users' secrecy requirements. In \cite{8365848}, a HLWNet-based security protocal was proposed for vehicular platoon communications. With this method, LiFi provides resilience to security attacks and WiFi offers redundancy for better reliability. In \cite{8481684}, the secrecy performance was analyzed for dual-hop HLWNets, where energy harvested from the optical signals of LiFi is used to relay data through RF. Similar work was reported in \cite{8606043}, with the aim to find the minimum transmission power that satisfies a certain secrecy rate.

\subsection{Challenges and Research Directions}

While HLWNets are able to benefit from combining the advantages of LiFi and WiFi, there are still some challenges yet to be overcome, in order to achieve a highly efficient network integration. For the application in indoor positioning, detecting 3-D positions in the HLWNet-based IPS has not yet been studied. How to efficiently utilize the different wireless technologies in HLWNets to locate device's positions, especially 3-D positions, is still an open issue. Detecting the device's orientation, which is required by many services including navigation, is another open issue to HLWNets. For the application in PLS, it is necessary to make a  trade off between the high security performance and the convenient system implementation in HLWNets. Currently the QKD has the unconditional security but this technology is not widely used due to the huge cost in implementation. On one hand, further development in miniaturization and feasibility will allow the massive deployment of low cost QKD devices in our daily-life scenario \cite{sibson2017chip,chun2017handheld}. On the other hand, since the wireless channel also has the inherent randomness and it is easier to implement the corresponding system, combination of the secure key generation in wireless channel and the quantum channel in HLWNets is an open issue.

\section{Conclusion} \label{sec:conclusion}

The inevitable trade-off between data transfer rate and coverage range encourages the coexistence of LiFi and WiFi, composing HLWNets which are a promising solution to future indoor wireless communications. HLWNets can not only greatly improve network performance but can also benefit application services such as indoor positioning and physical layer security. Handovers are identified as a key issue in HLWNets, especially a decision between a vertical handover and a horizontal handover. Low-complexity load balancing techniques are essential to HLWNets, due to the fact that WiFi APs are susceptible to traffic overload. Consequently, a central control unit, e.g. an SDN platform, is fundamental for an efficient use of HLWNets.





\section*{Acknowledgement}

This work was supported by the Engineering and Physical Sciences Research Council (EPSRC) grant EP/L020009/1: Towards Ultimate Convergence of All Networks (TOUCAN). Professor Harald Haas greatly acknowledges support from the EPSRC under Established Career Fellowship Grant EP/R007101/1.

\bibliographystyle{IEEEtran}
\bibliography{IEEEabrv.bib,HLWNet_Survey.bib}

\end{document}